\documentclass[preprint]{emulateapj} 

\usepackage{natbib}
\usepackage{hyperref}
\hypersetup{colorlinks,citecolor=Blue,linkcolor=Red,urlcolor=Blue}
\usepackage{amsmath}
\usepackage{empheq}
\usepackage[usenames,dvipsnames]{color}
\usepackage{enumitem}
\setlist{noitemsep}

\allowdisplaybreaks 

\newcommand{\appropto}{\mathrel{\vcenter{\offinterlineskip\halign{\hfil$##$\cr\propto\cr\noalign{\kern2pt}\sim\cr\noalign{\kern-2pt}}}}}

\newcommand{\Poincare}{{Poincar$\acute{\rm{e}}$}}
\newcommand{\Ham}{\mathcal{H}}
\newcommand{\G}{\mathcal{G}}

\newcommand{\p}{k}
\newcommand{\q}{\ell}

\newcommand{\Hsec}{\overline{\overline{\Ham}}_{\rm{s}}}

\newcommand{\HresM}{\Ham_{\rm{r}}}
\newcommand{\Hres}{\overline{\Ham}_{\rm{r}}}
\newcommand{\Hresec}{\overline{\Ham}_{\rm{rs}}}

\newcommand{\Msun}{M_{\odot}}

\newcommand{\phires}{\phi_{\rm{res}}}
\newcommand{\psires}{\psi_{\rm{res}}}
\newcommand{\varres}{\varphi_{\rm{res}}}

\begin{document}
 
\title{Dynamical Evolution Induced by Planet Nine}

\author{Konstantin Batygin$^1$, Alessandro Morbidelli$^2$} 
\affil{$^1$Division of Geological and Planetary Sciences, California Institute of Technology, 1200 E. California Blvd., Pasadena, CA 91125}
\affil{$^2$Laboratoire Lagrange, Université C\^ote d'Azur, Observatoire de la C\^ote d'Azur, CNRS, CS 34229, 06304 Nice, France}

\begin{abstract}
The observational census of trans-Neptunian objects with semi-major axes greater than $\sim250\,$AU exhibits unexpected orbital structure that is most readily attributed to gravitational perturbations induced by a yet-undetected, massive planet. Although the capacity of this planet to (i) reproduce the observed clustering of distant orbits in physical space, (ii) facilitate dynamical detachment of their perihelia from Neptune, and (iii) excite a population of long-period centaurs to extreme inclinations is well established through numerical experiments, a coherent theoretical description of the dynamical mechanisms responsible for these effects remains elusive. In this work, we characterize the dynamical processes at play, from semi-analytic grounds. We begin by considering a purely secular model of orbital evolution induced by Planet Nine, and show that it is at odds with the ensuing stability of distant objects. Instead, the long-term survival of the clustered population of long-period KBOs is enabled by a web of mean-motion resonances driven by Planet Nine. Then, by taking a compact-form approach to perturbation theory, we show that it is the secular dynamics embedded within these resonances that regulates the orbital confinement and perihelion detachment of distant Kuiper belt objects. Finally, we demonstrate that the onset of large-amplitude oscillations of orbital inclinations is accomplished through capture of low-inclination objects into a high-order secular resonance and identify the specific harmonic that drives the evolution. In light of the developed qualitative understanding of the governing dynamics, we offer an updated interpretation of the current observational dataset within the broader theoretical framework of the Planet Nine hypothesis. 
\end{abstract}

\section{Introduction}\label{sect1}

Over the course of one hundred and seventy years that followed the successful prediction of Neptune \citep{LeVerrier1846,Adams1846}, the presence of additional planets within the solar system has been contemplated by an extensive list of astronomers and celestial mechanicians \citep{Hoyt1980}. Historically, the lines of evidence for the existence of distant, massive bodies that orbit the sun have ranged from the (apparently) anomalous motion of Uranus \citep{1909AnHar..61..109P,1915MmLow...1....1L} and unexpected orbital characteristics of long-period comets \citep{Forbes1880,Whitmire1984}, to the peculiar structure of the solar system's small body populations \citep{2002Icar..160...32B,2006ApJ...643L.135G,2006Icar..184..589G,2008AJ....135.1161L,TS14,2017AJ....154...62V}. The predicted physical and orbital properties of the putative planets have been equally as varied, with inferred masses and semi-major axes spanning the Mars-Jupiter range and tens to thousands of astronomical units, respectively. 

\begin{figure}[t]
\centering
\includegraphics[width=\columnwidth]{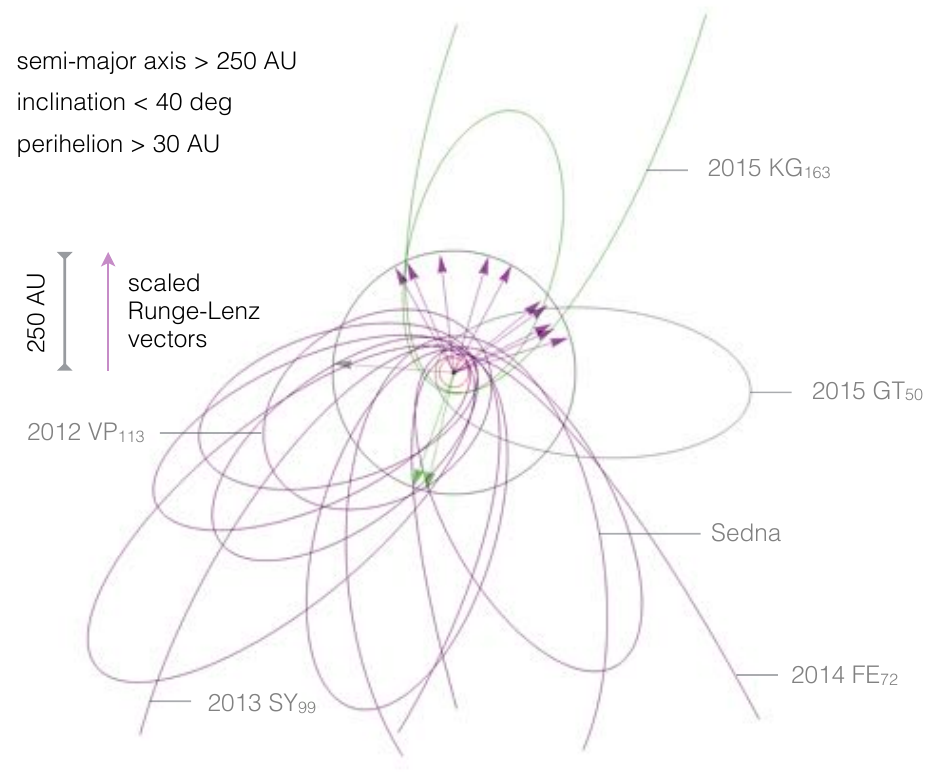}
\caption{Current observational census of the distant Kuiper belt. Thirteen known objects with semi-major axes greater than $250\,$AU and perihelion distance greater than $30\,$AU are shown in physical space, and are color-coded in accord with their dynamical class, as dictated by the Planet Nine hypothesis. Orbits belonging to the primary longitude of perihelion cluster (inferred to be apsidally anti-aligned with the orbit of Planet Nine) are shown in purple. Orbits that are diametrically opposed to the primary cluster (presumed to be apsidally aligned with the orbit of Planet Nine) are shown in green. The outlying object that does not correspond to either population is shown in gray. Each orbit is further labeled by its scaled Runge-Lenz vector, which is color-coded in the same way.}
\label{Fig:orbits}
\end{figure}

A recent addition to the aggregate of planetary predictions within the solar system is the Planet Nine hypothesis\footnote{The Planet Nine hypothesis was inspired by the work of \citet{TS14}, who noted that the arguments of perihelion (the angle between the apsidal and nodal lines on an orbit) of distant Kuiper belt objects are grouped together. In contrast with this finding, the primary aim of Planet Nine's inferred influence is to explain the simultaneous clustering of the longitudes of perihelion (a proxy for the direction of the pericenter in physical space) and the longitudes of ascending node (orientation of the orbital plane).} \citep{BB16}. Within the framework of this model, the observed orbital clustering of $a\gtrsim250\,$AU Kuiper belt objects (Figure \ref{Fig:orbits}) is sculpted by a $m\sim10\,m_{\oplus}$ planet residing on an appreciably eccentric ($e\sim0.3-0.7$), large semi-major axis ($a\sim 300-700\,$AU) orbit, whose plane roughly coincides with the plane of the distant bodies, and is characterized by perihelion direction that is anti-aligned with respect to the average apsidal orientation of the KBOs. In addition to (i) facilitating the orbital confinement of the aforementioned population of long-period KBOs and (ii) providing a physical mechanism for the perihelion detachment of Sedna-type orbits from Neptune, the presence of Planet Nine entails a series of additional consequences for the observed structure of the solar system \citep{BroBat}. In particular, it has been shown that the dynamical effects of Planet Nine naturally explain (iii) the existence of highly inclined, large semi-major axis Centaurs \citep{Gomes2015,BB16}, (iv) the six-degree obliquity of the sun \citep{2016AJ....152..126B,2016AJ....152..215L,2017AJ....153...27G}, as well as (v) the origins of proximate ($a<100\,$AU) retrograde KBOs \citep{BB16b}.

A key characteristic that differentiates the various planetary proposals is the dynamical mechanism through which the envisaged planet generates its observational signatures. In this regard, the Planet Nine (P9) hypothesis remains incomplete. While numerical simulations reveal that synthetic models of the solar system that include Planet Nine can provide a good match to the observational data (see however \citealt{2017arXiv170607447N}), the dynamical \textit{process} through which physical confinement of the distant orbits occurs remains poorly understood.

The original study of \citet{BB16} suggested that mean-motion resonances (including those of high order) are responsible for orbital clustering. Expanding on this idea, \citet{Malhotra2016} have suggested that each of the distant KBOs are currently trapped in $N$:1 and $N$:2 mean motion resonances (MMRs) with Planet Nine. More recently, \citet{ML2017,Becker2017} have carried out a large-scale numerical exploration of the resonant hypothesis in relation to the existing data, further demonstrating its viability. Meanwhile, \citet{Beust2016} has shown that a purely secular treatment of the dynamics provides a good match to the simulation results of \citet{BB16}, suggesting that resonant dynamics may be irrelevant to the problem at hand. In light of these conflicting results, the dominant P9-KBO interaction mechanism remains elusive, and its identification is the primary purpose of this paper.

Analytical characterization of P9-KBO coupling is important for three reasons, the first being falsifiability. The specific perturbation mechanism (along with the data it aims to explain) draws the distinction between various planetary hypotheses, and can be used to refute a specific model upon confrontation with the data. As an example, suppose that an observational survey discovers a planet at a radial separation of a few hundred AU, but the gravitational effects of this planet do not facilitate a physical confinement among the distant KBO orbits through the envisioned dynamical process\footnote{We note that the existence of a trans-Neptunian planet at an orbital separation of a few hundred AU was first proposed by \citet{Forbes1880}.}. Such a discovery would imply that the Planet Nine hypothesis, as formulated, is incorrect. 

A second, more practical motivation for studying P9-KBO interactions is the link between existing observations and the predicted orbit of Planet Nine. That is, if all distant KBOs are presently trapped in MMRs with P9, their mean longitudes may contain information about the location of Planet Nine on its orbit \citep{Malhotra2016,ML2017}. If, on the other hand, the interactions are purely secular, such a connection cannot be established\footnote{The process of orbital averaging inherent to secular perturbation theory removes all information related to the mean anomalies of the interacting bodies \citep{Morby2002}.}. Therefore, understanding the physics of P9-KBO coupling can offer a more complete interpretation of the extant observational dataset, and may yield an avenue towards further constraining the astronomical search for Planet Nine. 

The third, and final reason for this study is purely academic. While the exploration of the \textit{circular} restricted three-body problem (as a paradigm for interactions between planets and small bodies in the solar system) dates back multiple centuries \citep{Laplace,Poincare1902}, its highly elliptical counterpart remains much more scarcely understood (see e.g. \citealt{Beauge2006,Michtchenko2006,Gabrielle}). Indeed, a well-formulated theory for gravitational coupling between a planet and a test particle in the severe orbit-crossing regime does not currently exist. As a result, a perturbative study of P9 dynamics can yield interesting insights into the general mathematical structure of interactions among highly excited orbits. Such a framework would have considerable applications beyond the problem at hand, including the characterization of the remarkable dynamical states of highly eccentric, resonant exoplanets \citep{2004ApJ...611..517L,2013ApJ...777..101T}. 

Before delving into calculations, we delineate a list of specific questions we wish to answer in this work.
\begin{itemize}[topsep=1em]
\setlength\itemsep{0em}
\item What role (if any) do resonant interactions play within the dynamical evolution induced by Planet Nine? If resonances are prevalent, what order/multiplet harmonics dominate the dynamics, and what are their characteristic widths? \\
\item What role (if any) do secular interactions play within the dynamical evolution induced by Planet Nine? If dominant, how are close encounters avoided on nearly co-planar, anti-aligned orbits? Moreover, if resonant interactions are relevant to the Planet Nine hypothesis, why does the purely secular phase-space portrait provide a good match to the results of numerical simulations? \\
\item What parameters determine the critical semi-major axis corresponding to the transition between randomized and clustered longitudes of perihelion? What physical effect controls this transition? \\
\item What is the qualitative behavior of inclination dynamics within the framework of P9-driven evolution? What dynamical process allows some of the objects to acquire exceptionally high inclinations in the distant Kuiper belt?
\end{itemize}

\begin{figure*}[t]
\centering
\includegraphics[width=\textwidth]{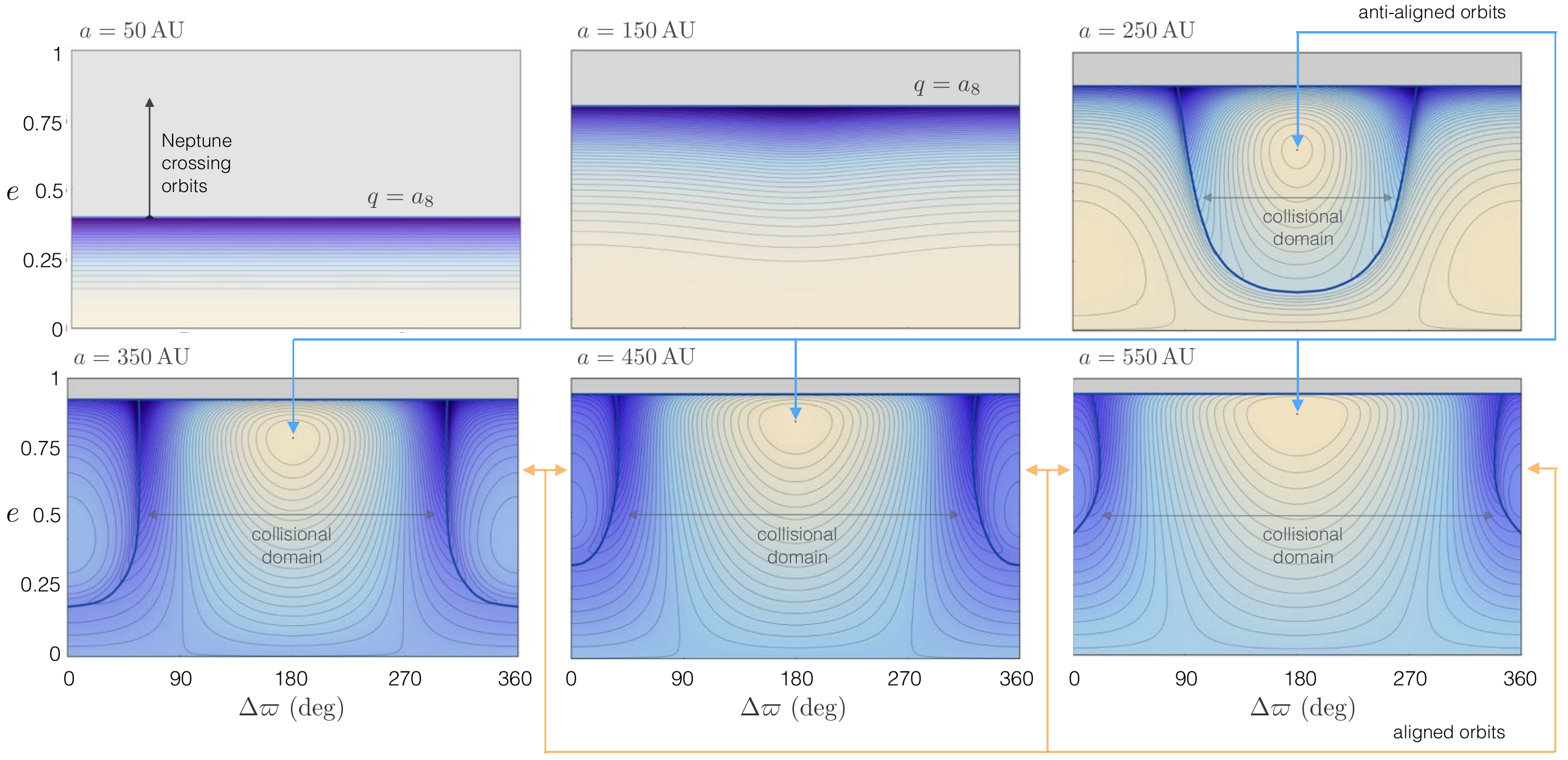}
\caption{Purely secular dynamics induced upon Kuiper belt objects by Planet Nine. Each panel is labeled by the corresponding value of the particle's semi-major axis, and depicts contours of the doubly averaged Hamiltonian (\ref{Hpuresec}) in eccentricity - longitude of perihelion (measured relative to the apsidal line of Planet Nine) space. In panels characterized by $a\geqslant250\,$AU, secular equilibria corresponding to both apsidally aligned ($\Delta\varpi=0$) as well as anti-aligned ($\Delta\varpi=180\deg$) configurations emerge, and are segregated by tangential collision curves, shown as thick curves. Note that in spite of the apparent smoothness of the level curves of the averaged Hamiltonian surrounding the anti-aligned equilibrium points, this region of phase-space describes orbital configurations that are not protected from close encounters, and thus entails a long-term unstable orbital evolution. Moreover, notice that only apsidally aligned configurations that are close to the $\Delta\varpi=0$ equilibrium points are protected from close encounters by the geometrical collinearity of the orbits.}
\label{Fig:secular}
\end{figure*}

We take these questions as an approximate guide to the logic of the paper, which is structured as follows. In section \ref{sect2}, we revisit a purely secular description of P9-induced orbital evoluition, carrying out the averaging procedure in closed form. In section \ref{sect3}, we consider a pair of idealized $N$-body simulations and outline the key differences between numerical experiments and pure secular theory. In section \ref{sect4}, we present a semi-analytical theory of resonant P9-KBO interactions, and elucidate secular dynamics embedded within mean-motion resonances as the primary driver of apsidal confinement of distant KBOs. Subsequently, we discuss the onset of large-scale inclination oscillations of long-period bodies in section \ref{sect5}. We then re-examine the existing observational data in light of the aforementioned theoretical developments in section \ref{sectobs}. We summarize and discuss the implications of our results in section \ref{sect6}. A brief analysis of long-term stable, but apsidally unconfined orbits is presented in the appendix.

\section{Purely Secular Dynamics}\label{sect2}

We begin our study of Planet Nine-induced dynamics within the framework of purely secular perturbation theory. In \citet{BB16}, our preliminary exploration of secular dynamics relied on an octupole order expansion of the gravitational potential \citep{1962AJ.....67..300K,2013MNRAS.435.2187M}, that implicitly assumed that the orbits under consideration do not cross. Following \citet{Beust2016}, here we abandon the series-expansion approach to modeling P9-KBO coupling, and carry out the phase-averaging procedure in closed form. 

As in \citet{BB16,Beust2016}, we assume coplanarity, and model the mean-field effects of the known giant planets neglecting their eccentricities. Further, we choose to work in a slowly rotating coordinate frame, that co-precesses with Planet Nine's perihelion (the corresponding contact transformation is spelled out in \citealt{BB16}). The governing (doubly averaged) Hamiltonian of a test particle under planetary perturbations then has the form:
\begin{align}
\Hsec &= -\frac{1}{4}\frac{\G\,\Msun}{a}\frac{1}{\big(1-e^2\big)^{3/2}} \sum_{j=5}^{8}\frac{m_j\,a_j^2}{\Msun\,a^2} \nonumber \\
&+\dot{\varpi}_9 \,\sqrt{\G\,\Msun a}\,\big(1-\sqrt{1-e^2} \big) \nonumber \\
&- \frac{1}{4\pi^2}\oint \oint \frac{\G\,m_9}{|\mathbf{r}-\mathbf{r}_9|} \,d\lambda\,d\lambda_9,
\label{Hpuresec}
\end{align}
where $\G$ is the gravitational constant, $\Msun$ is the mass of the sun, $\mathbf{r}$ is the position vector, $\lambda$ is the mean longitude, $a$ is semi-major axis, $e$ is eccentricity, and $\varpi$ is the longitude of perihelion. All quantities pertaining to the four canonical giant planets are labeled with indexes $5-8$, while the values corresponding to Planet Nine are denoted with the subscript 9. The unlabeled variables correspond to the Kuiper belt object. Finally, Planet Nine's orbit-averaged precession rate is given by the expression\footnote{Note that there is a typo in the corresponding expression in \citet{BB16}.}:
\begin{align}
\dot{\varpi}_9 = \frac{3}{4}\sqrt{\frac{\G\,\Msun}{a_9^3}}\frac{1}{\big(1-e_9^2\big)^{2}} \sum_{j=5}^{8}\frac{m_j\,a_j^2}{\Msun\,a_9^2}.
\label{P9precc}
\end{align}

The three terms present in equation (\ref{Hpuresec}) have simple physical interpretations. The first term governs the secular advance of the KBO's perihelion due to the phase-averaged gravitational potential of Jupiter, Saturn, Uranus, and Neptune (in direct parallel with equation \ref{P9precc}). The second term accounts for the fact that the reference frame is slowly rotating. The third term governs secular P9-KBO interactions. Note that the indirect part of the disturbing potential is by default entirely averaged out within the framework of secular theory, and need not be accounted for \citep{Morby2002}.

The only pair of dynamical variables present in Hamiltonian (\ref{Hpuresec}) is ($e,\Delta\varpi$), meaning that the system is integrable. In other words, contours of the numerically averaged function (\ref{Hpuresec}) fully encapsulate the accompanying orbital evolution. Figure (\ref{Fig:secular}) shows the secular phase-space portraits of the system, projected into $e-\Delta\varpi$ space, for $a=50,150,...,550\,$AU, adopting P9 parameters from \citet{BB16} (specifically, $a_9=700\,$AU, $e_9=0.6$ and $m_9=10\,m_{\oplus}$). This figure can be readily compared with Figure (4) of \citet{BB16}, and confirms that the purely secular portrait provides a good match to the numerically computed portraits in the same dynamical regime \citep{Beust2016}. 

On all panels denoted by $a\geqslant250\,$AU, $e-\Delta\varpi$ space is characterized by two stable equilibrium points: one at $\Delta\varpi=0$ and another at $\Delta\varpi=180\deg$. On each diagram, the two libration regions surrounding these fixed points are separated by a solid curve, which corresponds to a tangential configuration of the KBO and P9 orbits. Note further that on these tangential contact curves, the derivatives of the Hamiltonian are discontinuous (i.e. $\Hsec$ is locally class $\mathcal{C}^0$), signaling a breakdown of the secular framework \citep{2002CeMDA..83...97G}. 

A simple examination of the contours of $\Hsec$ depicted on Figure (\ref{Fig:secular}) reveals that KBOs initialized on nearly Neptune-crossing orbits (immediately below horizontal lines labeled by $q=a_8$) will suffer drastically different secular evolutions depending on their starting value of $\Delta \varpi$. Bodies initially close to $\Delta\varpi\sim0$ will be driven onto the tangential orbit-crossing curves through apsidal precession, and will eventually be removed from the system by recurrent close encounters with Planet Nine. On the other hand, objects initialized at $\Delta\varpi\sim\pi$ settle onto secular trajectories that encircle that anti-aligned equilibrium, and never encounter the tangential collision curves. Thus, the dynamical lifetimes of apsidally anti-aligned objects can be na\"{\i}vely envisioned to be longer than their aligned counterparts, and in time, a cluster of exclusively anti-aligned orbits should be carved out by Planet Nine.

\begin{figure*}[t]
\centering
\includegraphics[width=0.8\textwidth]{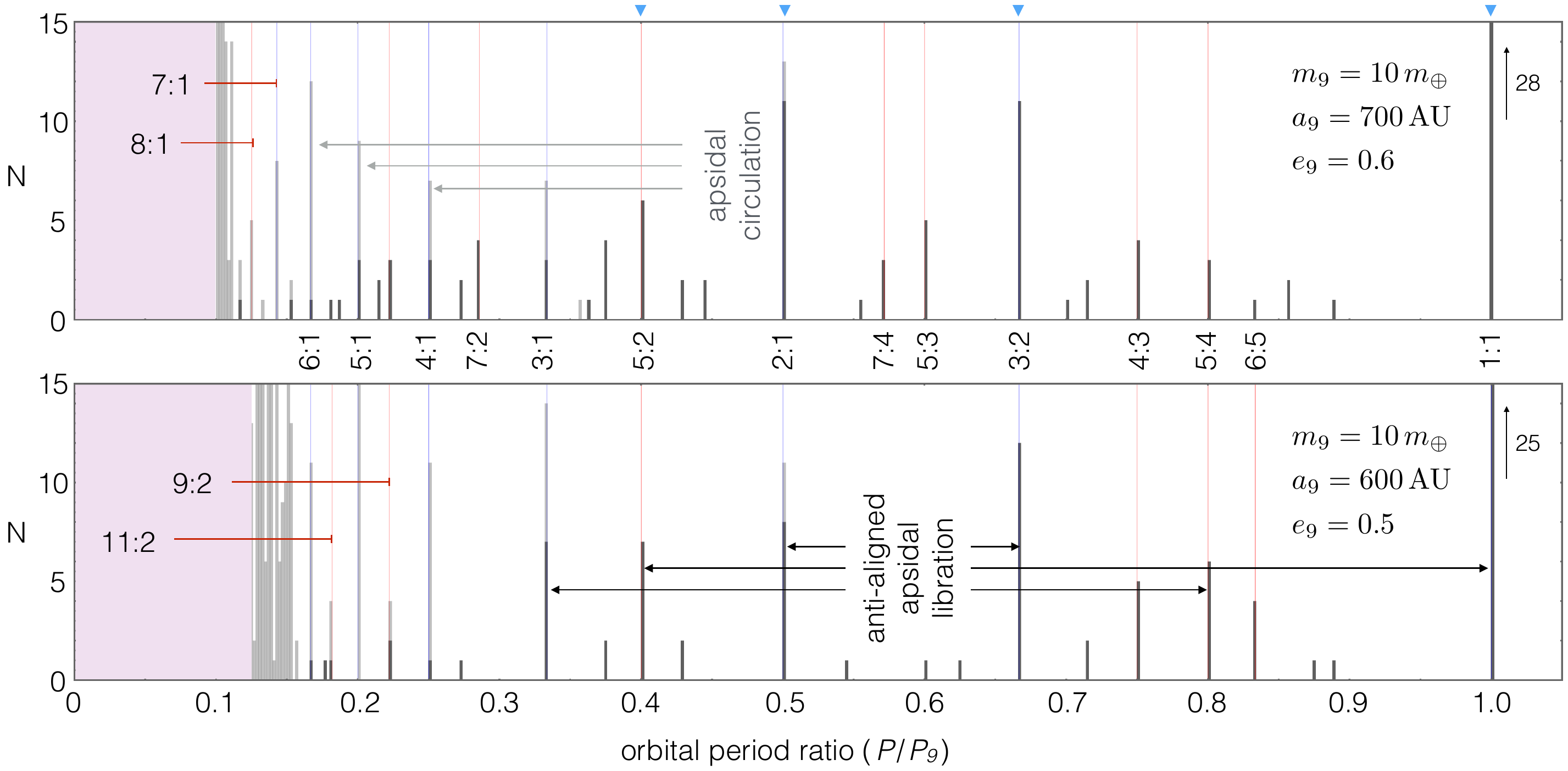}
\caption{A histogram summarizing the orbital distribution of simulated particles with dynamical lifetimes that exceed $4\,$Gyr. Beyond a critical orbital period ratio ($P/P_9\gtrsim0.1$ for $a_9=700\,$AU, $e_9=0.6$ and $P/P_9\gtrsim0.15$ for $a_9=600\,$AU, $e_9=0.5$), all surviving members of the distant scattered disk reside in mean-motion resonances with Planet Nine, and derive their long-term stability from the associated phase-protection mechanism. The final orbital configurations of distant bodies sculpted by Planet Nine are similar in both numerical experiments, and show the onset of clustering in longitude of perihelion beyond the 3:1 mean motion resonance. Correspondingly, bins containing particles locked in a stable pattern of anti-aligned apsidal libration are shown in black, while bins containing objects with circulating longitudes of perihelion are shown in gray. The commensurabilities possessing the largest number of apsidally confined particles are indicated with blue triangles.}
\label{Fig:histogram}
\end{figure*}

We note further that there is an island of stable apsidal libration around $\Delta \varpi=0$ that avoids crossing the tangential configuration curve and is thus protected from close-encounters. However, the eccentricities along these protected orbits are moderate and the corresponding perihelion distances are large. This means that even if KBOs somehow came to occupy these islands of stability, they would be difficult to detect observationally.

With this picture in mind, it is tempting to affirm that the agreement between theory and simulation is satisfactory, and proceed forward within the purely secular framework. This is however a misconception, facilitated by the apparent smoothness of the secular phase-space portraits shown in Figure (\ref{Fig:secular}). In particular, the fact that the contours of $\Hsec$ do not show any kinks within the apsidally anti-aligned domain is simply a consequence of the integrability of the non-tangential singularity \citep{1996CeMDA..64..209T,Gronchi1998}, and does not mean that the system can elude collisions. Instead, recalling that the physical setup of the orbits is planar, it is trivial to demonstrate that under the assumption of uncorrelated Keplerian motion, all objects entrained in the apsidally anti-aligned configuration with Planet Nine would suffer close-encounters on timescales much shorter than the age of the solar system. Therefore, despite giving the illusion of agreement with $N$-body simulations, pure secular theory predicts that the entire distant Kuiper belt is dynamically unstable, and should have been cleared out on a timescale comparable to the orbital precession time.

\begin{figure*}[t]
\centering
\includegraphics[width=1\textwidth]{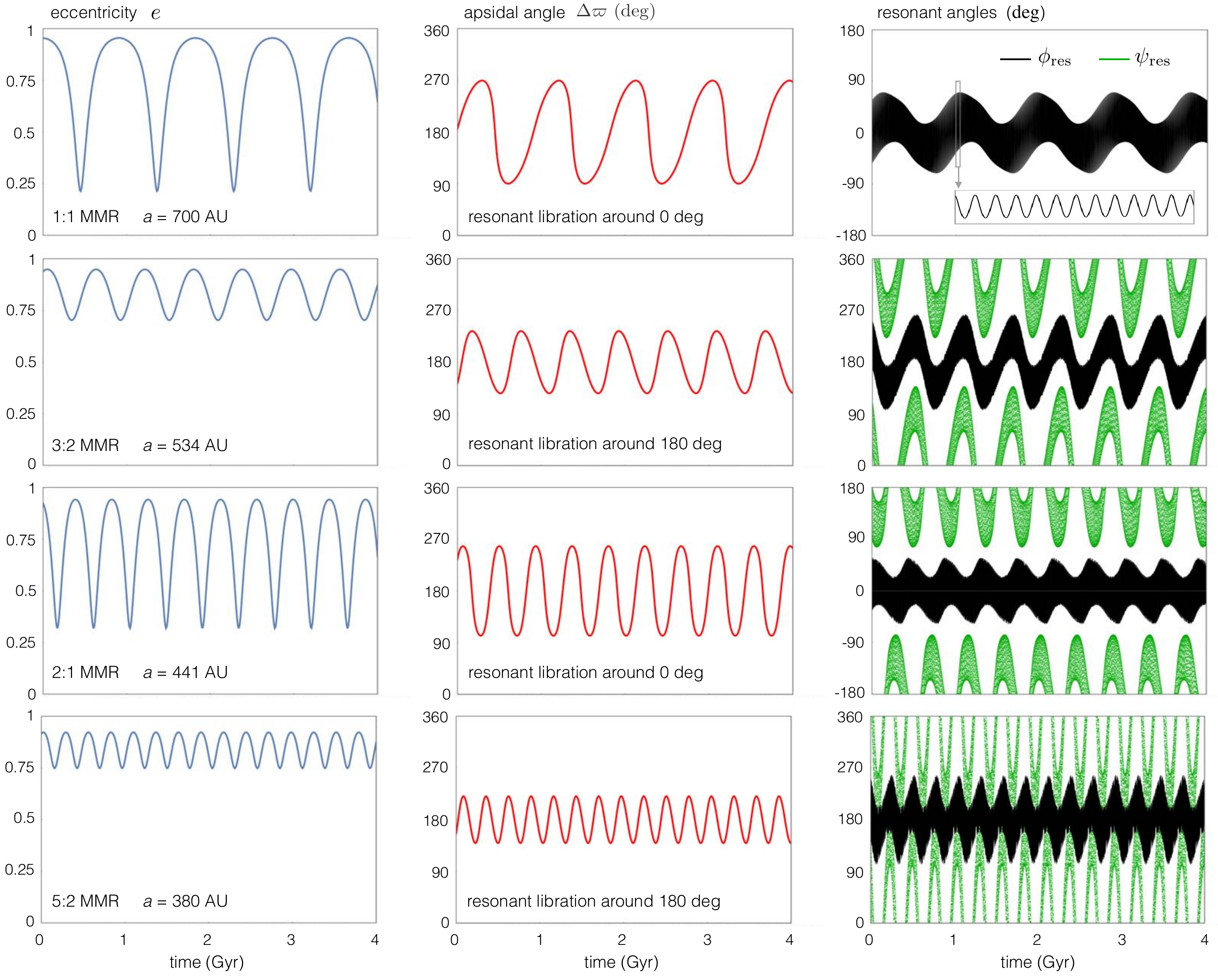}
\caption{Orbital time series of four resonant objects drawn from the $N$-body simulation with $m_9=10\,m_{\oplus}$, $a_9=700\,$AU and $e_9=0.6$. The left and middle panels show oscillations of the eccentricities and the relative longitudes of perihelion respectively, while the right panels depict the behavior of the resonant angles $\phires$ (black) and $\psires$ (green). Generally, particles that exhibit stable anti-aligned apsidal libration (such as those shown in this Figure) are characterized by low-amplitude oscillation of $\phires$ with a period equal to that of libration of $\Delta\varpi$ (not to be confused with the actual libration amplitude of $\phires$, which is short periodic and corresponds to the thickness of the black ``band", as shown in the inset of the top-right panel). On the contrary, orbital evolution driven primarily by low-amplitude libration of a resonant harmonic that contains $\varpi$ (rather than only $\varpi_9$ as in $\phires$) corresponds to circulation of the longitude of perihelion (examples shown in the appendix). Note that $\phires$ librates around 0 for the 1:1 \& 2:1 MMRs and around $180\deg$ for the 3:2 \& 5:2 MMRs.}
\label{Fig:timeseries}
\end{figure*}

Contrary to this view, published numerical experiments reveal that particles residing deep within the cores of the anti-aligned libration regions of Figure (\ref{Fig:secular}) remain stable over the multi-Gyr lifetime of the solar system \citep{BB16,BroBat}. This disparity suggests that secular theory alone is unlikely to represent a full dynamical description of P9-driven evolution, and an additional stabilizing mechanism is at play in the simulations. Let us now examine this point further. 

\section{Numerical Simulations}\label{sect3}
To quantify the discrepancy between published simulations and pure secular theory, we carry out a pair of simplified numerical experiments that mirror those reported in \citep{BB16,BroBat}. In particular, we evolve an initially axisymmetric disk of 6,000 eccentric test-particles with $a\in(150,750)\,$AU and $q\in(30,36)\,$AU, under the influence of the phase-averaged potential of the four inner giants as well as Planet Nine with $(a_9,e_9)=(700\,\rm{AU},0.6)$ and $(a_9,e_9)=(600\,\rm{AU},0.5)$. The test particles are initialized with a null vertical velocity dispersion and thus remain confined to Planet Nine's orbital plane throughout the simulation.

Unlike the simulation suite of \citet{BB16,BroBat,ML2017,Becker2017} where Neptune was modeled directly, or that of \citet{BB16b} where the Keplerian motion of all four inner giants was resolved, here we emulate the effects of Jupiter, Saturn, Uranus and Neptune with an effective quadrupolar gravitational moment of the Sun: 
\begin{align}
J_{2} = \bigg( \frac{3\cos^2(i_9)-1}{4} \bigg) \sum_{j=5}^{8} \frac{m_j\,a_j^2}{\Msun\,\mathcal{R}^2},
\label{Jeff}
\end{align}
setting the inner absorbing radius to $\mathcal{R}=20\,$AU. This idealization is employed specifically to avoid contaminating P9-induced KBO evolution with chaotic dynamics that arise from scattering off of Neptune, and to yield the closest point of comparison between numerical and semi-analytical results. Moreover, here we will assume that the inclination of Planet Nine relative to the Laplace plane of the canonical giant planets is sufficiently small to approximate $\cos(i_9)\approx1$, while keeping in mind that $\hat{z}-$axis of our coordinate system coincides with the orbital plane of Planet Nine.

To carry out the simulations, we utilized the \texttt{mercury6} gravitational dynamics software package \citep{1999MNRAS.304..793C}. The integrations were performed using the hybrid Wisdom-Holman/Bulirsch-Stoer algorithm \citep{1992AJ....104.2022W,1992nrca.book.....P}, with a time-step of $\Delta t=3,100\,$days (i.e. $1/10\,$th of Uranus' orbital period), and spanned $4\,$Gyr. Any particle that attained a radial distance of $r<\mathcal{R}$ or $r>10,000\,$AU was removed from the simulations. 

Within the context of these idealized numerical experiments, all long-term stable particles retain nearly constant semi-major axes throughout the integration, and Figure (\ref{Fig:histogram}) depicts an orbital histogram of the surviving bodies. Bins shown in black correspond to apsidally confined objects (defined by $|\Delta\varpi-180\deg|\leqslant 90\deg$ throughout the simulation), while those shown in gray denote particles that are long-term stable, but experience apsidal circulation. We note that here, the gray bins are stacked on top of the black bins, such that the relative size of gray and black components of each column is a measure of the contamination of apsidally confined population of objects (at a given semi-major axis) by those undergoing perihelion circulation.

 A strong propensity towards apsidal clustering for $P/P_9\gtrsim1/3$ is clearly evident in both simulations. However, it is also important to note that the same orbital periods (e.g. those corresponding to the 3:1, 2:1, and 1:1 commensurabilities in Figure \ref{Fig:histogram}) can be simultaneously occupied by apsidally circulating and librating orbits. This means that even within the framework of highly idealized treatment of P9-induced dynamics, the clustering of the longitude of perihelion beyond a critical semi-major axis is not perfectly strict, and the existence of objects that do not conform to the general anti-aligned orbital pattern is an expected consequence of the model \citep{Shankman}.

In addition to the already established tendency towards orbital clustering with increasing orbital period, it can be clearly seen in Figure (\ref{Fig:histogram}) that all long-lived objects which exhibit perihelion confinement have semi-major axes that correspond to mean-motion commensurabilities with Planet Nine. Particularly, the 1:1, 3:2, 2:1, and 5:2 resonances contain the largest populations of apsidally clustered KBOs in both simulations\footnote{We note that these simulations are not intended to represent a full exploration of the resonant capture probabilities within the context of the Planet Nine hypothesis. A more complete estimation these probabilities is presented in a companion paper \citep{Baileyinprep}.}. This point undercuts a key difference between the results of $N$-body simulations and pure secular theory, and demonstrates that (at least within the framework of a planar physical setup) distant KBOs derive their long-term orbital stability from the phase-protection mechanism inherent to mean-motion resonances. 

The typical dynamical evolutions of bodies trapped in the four aforementioned resonances are shown in Figure (\ref{Fig:timeseries}). Specifically, the left and middle panels depict the evolutions of orbital eccentricities and apsidal lines. On timescales of order $\sim 0.1-1\,$Gyr, orbital eccentricities experience considerable oscillations in concert with librations of $\Delta\varpi$. This facilitates a periodic regression of the KBO perihelion distance, generating dynamically detached (Sedna-type) orbits \citep{2002Icar..157..269G,2004ApJ...617..645B}. 

Of particular importance is the behavior of the resonant angles:
\begin{align}
\psires &= \p\,\lambda_9-\q\,\lambda-(\p-\q)\,\varpi, \nonumber \\
\phires &= \p\,\lambda_9-\q\,\lambda-(\p-\q)\,\varpi_9,
\label{phi}
\end{align}
where $\p$ and $\q$ are integers (note that for $|\p-\q|>1$, additional resonant harmonics related to these angles through $\Delta\varpi$ exist). On the right panels of Figure (\ref{Fig:timeseries}), $\psires$ is plotted in green and $\phires$ is plotted in black. The actual librations of $\phires$ and $\psires$ are short-periodic (as portrayed by the inset within the top-right panel of Figure \ref{Fig:timeseries}) and their amplitudes correspond to the thickness of the green and black ``bands."  Meanwhile, the long-periodic oscillations of these angles are mere reflections of the librations of $\Delta\varpi$, which modulate the locations of the resonant equilibria associated with $\phires$ and $\psires$. 

The fact that the amplitude of long-periodic oscillations of $\phires$ is much smaller than that of $\psires$ for orbits that exhibit libration in $\Delta\varpi$, demonstrates that the resonant multiplets containing $\phires$ represent a better approximation of the real dynamics than those containing $\psires$. In particular, if the Hamiltonian depended only on $\phires$, this angle would have no long-periodic oscillations while $\psires$ (being equal to $\phires + (\ell-k)\,\Delta\varpi$) would oscillate with the amplitude and period of $|(\ell-k)|\,\Delta\varpi$. The opposite would be true if the Hamiltonian dependent only on $\psires$. Thus the relative amplitudes of long-periodic oscillations of the these angles are inversely correlated to the relative strengths of the corresponding terms. Accordingly, in our analytic approach in the next section we will consider $\phires$ as a reference angle, in contrast to \citet{Malhotra2016} and \citet{Beust2016}, who considered $\psires$ instead. 

Only containing the longitude of perihelion of Planet Nine and not that of the Kuiper belt objects themselves, the (short-periodic) libration of $\phires$ drives the oscillation of the particle's semi-major axis but does not affect the evolution of its eccentricity. Thus, the libration in $\phires$, sometimes referred to as ``corotation" resonance, already implies a certain disconnect among the degrees of freedom related to the particle's semi-major axis and its eccentricity. Moreover, the striking separation of timescales associated with resonant ($\phires-a$) dynamics and secular ($e-\Delta\varpi$) dynamics motivates the construction of the semi-analytical model that will follow, based on the adiabatic approach.

The dominant dependence of the resonant dynamics on $\phires$, as opposed to another harmonic that contains $\varpi$ in its critical argument, is central to maintaining apsidally anti-aligned libration of the orbits. In fact, long-term stable particles whose resonant dynamics is driven by any angle other than $\phires$ exhibit circulation of the longitude of perihelion, and correspond to objects denoted as gray bins in Figure (\ref{Fig:histogram}). Although characterizing the dynamics in this transitionary semi-major axis range is not the primary purpose of this study, we present a brief analysis of this mode of orbital evolution in the appendix. 

\section{Semi-Analytical Theory} \label{sect4}

\begin{figure}[t]
\centering
\includegraphics[width=0.95\columnwidth]{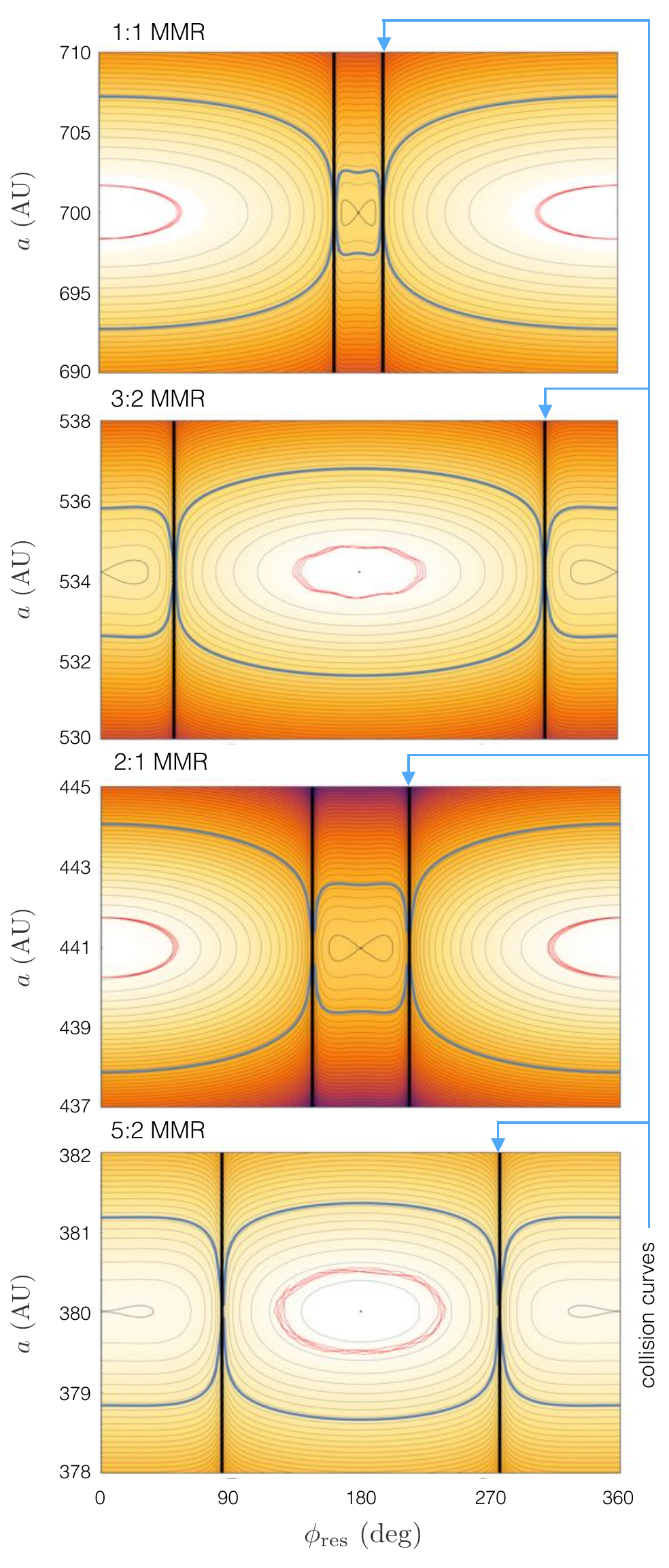}
\caption{Resonant phase-space portraits of the four most populated mean-motion commensurabilities, projected into $a-\phires$ space. The color scale and the black contours correspond to level curves of the singly averaged Hamiltonian (\ref{Hres}), while the vertical black lines denote collision curves. Resonant contours of the Hamiltonian that come into contact with the collision curves are shown as thick purple lines, and thereby inform the widths of the corresponding resonances. These phase-space diagrams adopt our fiducial P9 parameters and assume that the Kuiper belt objects are characterized by a fixed value of $\Gamma$ that corresponds to $q=35\,$AU at nominal resonance and $\Delta\gamma=\pi$. The red curves that encircle the elliptic equilibria in each panel represent the $a-\phires$ evolution of the particles shown in Figure (\ref{Fig:histogram}) near $\Delta\varpi=\pi$, and signal excellent agreement with theory.}
\label{Fig:resonant}
\end{figure}

As shown in the previous section, all surviving members of the synthetic scattered disk that exhibit persistent apsidal anti-alignment with Planet Nine's orbit are locked into mean-motion resonances with P9. The purpose of the following analysis is thus to explain this behavior from semi-analytic grounds. To achieve this goal, we consider the isolated resonant behavior first. As in section (\ref{sect2}), we will abandon traditional methods of celestial mechanics based on series-expansions, and cast our analysis of the governing Hamiltonian in closed form.

\subsection{Mean Motion Resonances}\label{sect41}

In order to carry out the averaging process numerically and elucidate the $\phires-a$ resonant dynamics, we set up a rigorous Hamiltonian approach to the problem. As is typical for the planar restricted three body problem, the canonical \Poincare\ action-angle variables for the test particle are \citep{MD99}:
\begin{align}
&\Lambda=\sqrt{\G \Msun \, a}     &\lambda=\mathcal{M}+\varpi \nonumber \\
&\Gamma=\sqrt{\G \Msun \, a}\big(1-\sqrt{1-e^2} \big)   &\gamma=-\varpi
\label{Poincvar}
\end{align}
where $\mathcal{M}$ is the mean anomaly. In terms of these variables, the Hamiltonian of the problem is \citep{Morby2002}
\begin{align}
\HresM = -\frac{1}{2} \bigg( \frac{\G\,\Msun}{\Lambda} \bigg)^2 - \G\,m_9 \, \bigg(\frac{1}{|\mathbf{r}-\mathbf{r}_9|} - \frac{\mathbf{r}\cdot \mathbf{r}_9}{|\mathbf{r}_9 |^3} \bigg).
\label{Hammyres}
\end{align}

\begin{figure*}[t]
\centering
\includegraphics[width=1\textwidth]{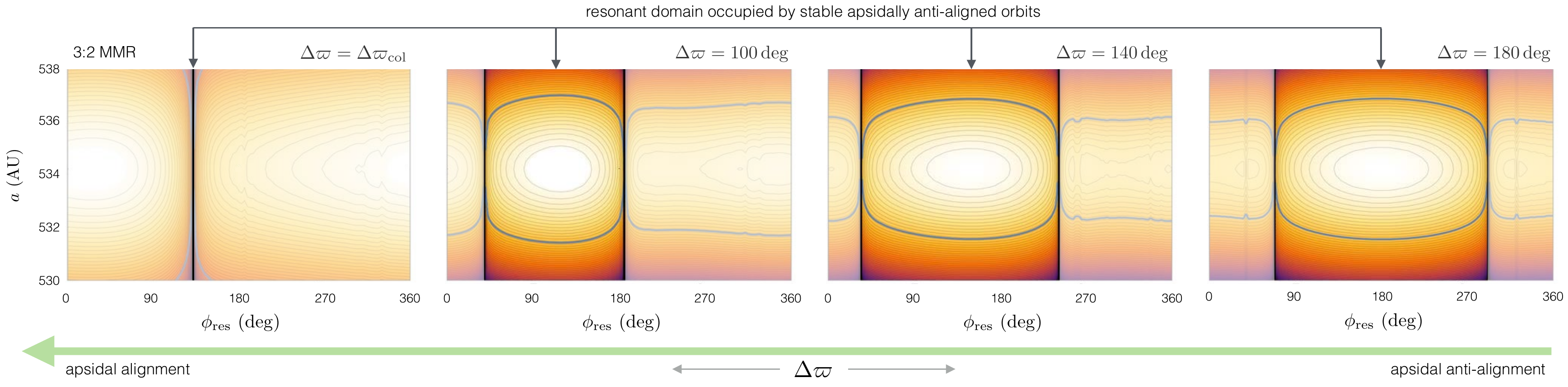}
\caption{Resonant phase-space portraits associated with the 3:2 mean motion commensurability at different values of the relative longitude of perihelion, $\Delta\varpi$. As the orbital configuration progressively shifts away from exact anti-alignment of perihelia, the resonant domain occupied by stable orbits shown in Figure (\ref{Fig:resonant}) shrinks, until it is fully engulfed by the collision curves. Correspondingly, this process imposes a limit on maximal deviation from strict apsidal anti-alignment that a resonant trajectory can endure before experiencing close encounters with Planet Nine. The depicted portraits were computed assuming a value of $\Gamma$ that corresponds to the equilibrium eccentricity of the resonant-secular phase-space portrait, and their ($e,\Delta\varpi$) coordinates are shown with open circles on Figure (\ref{Fig:secres}).}
\label{Fig:varpimod}
\end{figure*}

The vector $\mathbf{r}$ can be written as a function of the canonical variables (\ref{Poincvar}). Similarly, $\mathbf{r}_9$ can be expressed as a function of the parameters $a_9, e_9$, which are assumed to be fixed in time, as well as the angles $\lambda_9$ and $\varpi_9$, which are assumed to advance with fixed frequencies, $n_9$ and $\dot{\varpi}_9$ respectively. Thus, Hamiltonian (\ref{Hammyres}) depends on time through these two planetary angles.

In order to remove explicit time dependence from equation (\ref{Hammyres}), we extend the phase space by two degrees of freedom. That is, we consider $\lambda_9$ and $\gamma_9=-\varpi_9$ as independent variables, with conjugated actions $\Lambda_9$ and $\Gamma_9$. Correspondingly, we add the term $(n_9\, \Lambda_9 - \dot{\varpi}_9 \, \Gamma_9)$ to expression (\ref{Hammyres}), making the Hamiltonian autonomous. Of course, in doing so we do not alter the dynamics in any way, since $d\lambda_9/dt=\partial \HresM/\partial \Lambda_9=n_9$ and $d\varpi_9/dt = -d\gamma_9/dt = -\partial{\HresM}/\partial \Gamma_9 =\dot{\varpi}_9$.

We now denote the mean motion of the particle by $n=\G^2\,\Msun^2/\Lambda^3$ and assume that there is a resonance of the kind $\p\,n_9 - \q\,n=0$ for some integers $\p$ and $\q$. It is then appropriate to make the following canonical transformation\footnote{This contact transformation arises from a type-2 generating function of the form $\mathcal{F}_2=-\Lambda' (\p\lambda_9-\q\lambda+(\p-\q)\gamma_9)+\Lambda'_9(\lambda_9)+\Gamma'(\gamma-\gamma_9)+\Gamma'_9(\gamma_9)$.}
\begin{align}
&\Lambda'= \Lambda/\q & \phires'=-\phires \nonumber \\
&\Lambda'_9= \Lambda_9+(\p/\q)\, \Lambda  &\lambda'_9=\lambda_9  \nonumber \\
&\Gamma'=\Gamma &\Delta \gamma=\gamma-\gamma_9 = -\Delta\varpi \nonumber \\
&\Gamma'_9=\Gamma_9+\Gamma+((\p-\q)/\q)\,\Lambda  &\gamma'_9=\gamma_9.
\label{transvar}
\end{align}

\begin{figure*}[t]
\centering
\includegraphics[width=0.8\textwidth]{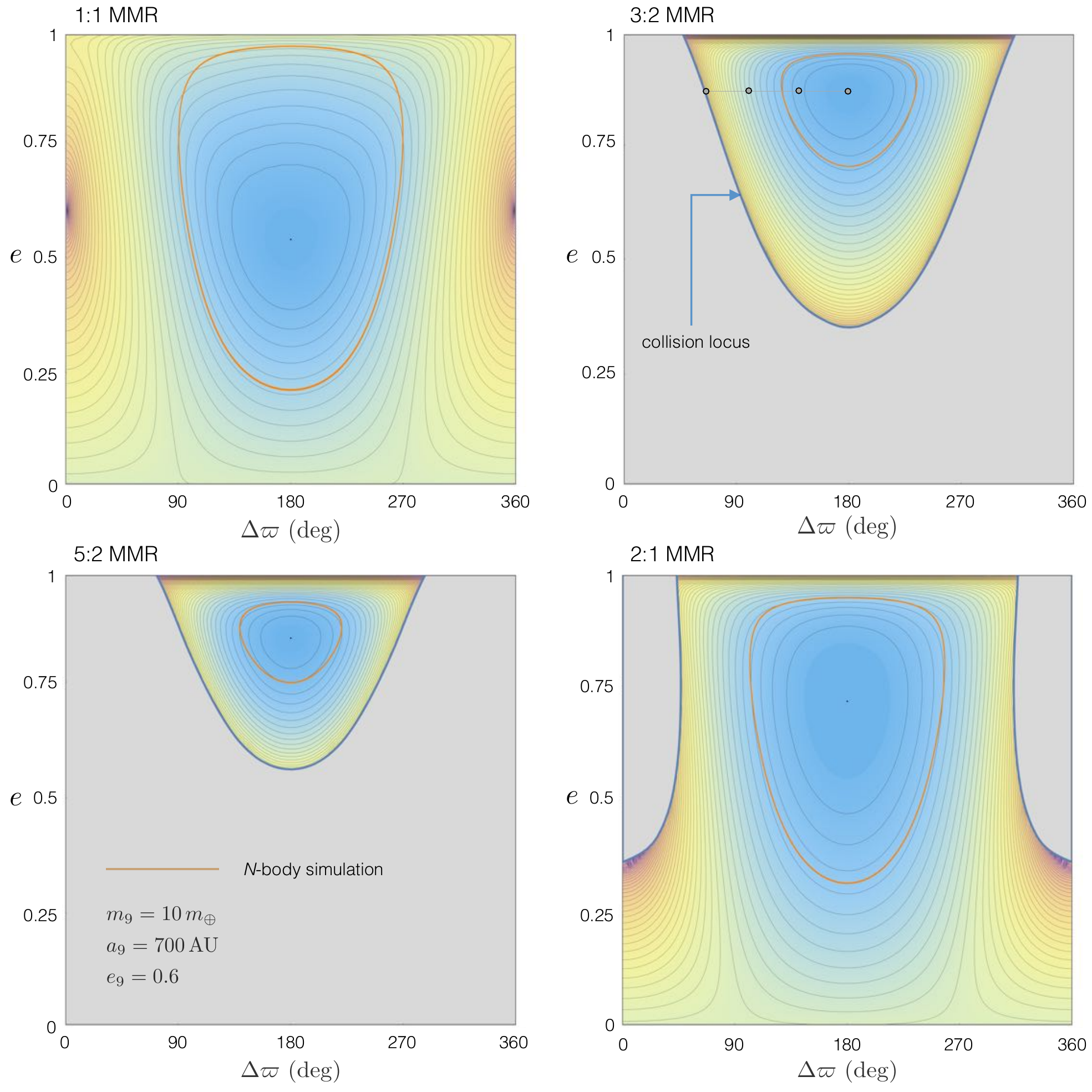}
\caption{Eccentricity-perihelion diagrams showing long-term stable secular evolution facilitated by phase protection arising from $\phires$-type resonances. For resonances other than 1:1, the $e-\Delta\varpi$ domain is restricted by a collision locus, which arises from engulfment of the stable resonant equilibrium shown in Figure (\ref{Fig:resonant}) by collision curves. Contours of the singly averaged resonant-secular Hamiltonian (\ref{Hsecres}) are shown as black lines that follow the background color-scale, and $N$-body trajectories of particles shown in Figure (\ref{Fig:timeseries}) are over-plotted as orange curves. The clear agreement between semi-analytic perturbation theory and numerical experiments demonstrates that although the long-term survival of distant Kuiper belt objects is enabled by resonant interactions, the clustering of longitudes of perihelion and dynamical detachment of orbits from Neptune arise from secular perturbations embedded within the resonances. The ($e,\Delta\varpi$) coordinates of resonant phase-space portraits depicted in Figure (\ref{Fig:varpimod}) are shown with open circles on the panel corresponding to the 3:2 MMR.}
\label{Fig:secres}
\end{figure*}

The unperturbed part of the Hamiltonian (the term independent of $m_9$) becomes:
\begin{align}
\big(\HresM\big)_{\rm{kep}} &= -\frac{1}{2} \bigg( \frac{\G\,\Msun}{\Lambda} \bigg)^2 +n_9(\Lambda'_9-\p\, \Lambda') \nonumber \\
&-\dot{\varpi}_9\,(\Gamma'_9-\Gamma-(\p-\q)\,\Lambda').
\label{Hammyreskep}
\end{align}
Because $\partial (\HresM)_{\rm{kep}} /\partial \Lambda'= \q\,n- \p\,n_9=0$ there is only one fast angle in the Hamiltonian, and it is $\lambda'_9$. Thus, the perturbation can be averaged over this angle, leading to
\begin{align}
\big(\Hres\big)_{\rm{pert}} &= -\frac{\G\,m_9}{2\,\pi \,\q} \int_0^{2\pi \q} \bigg(\frac{1}{|\mathbf{r}-\mathbf{r}_9|} - \frac{\mathbf{r}\cdot \mathbf{r}_9}{|\mathbf{r}_9 |^3} \bigg)\, d \lambda_9.
\label{Hammyrespert}
\end{align}

It is easy to see from d'Alembert rules, that the averaged Hamiltonian can depend only on two angles: $\phires'$ and $\Delta \gamma$. Thus, $\Lambda'_9$ and $\Gamma'_9$ are constants of motion and can be dropped. The Hamiltonian now comprises a two degree of freedom system, and is not integrable. However, we note that the degree of freedom in ($\Gamma',\Delta \gamma$) is characterized by a frequency of order $\propto m_9/\Msun$, and is slow relative to the other degree of freedom in ($\Lambda',\phires'$), whose frequency is of the order of $\propto\sqrt{m_9/\Msun}$ \citep{1990CeMDA..47...99H}. 

Taking advantage of the aforementioned separation of timescales between the resonant and secular dynamics, we may evaluate the phase-space portrait associated with the resonant Hamiltonian in the adiabatic approximation, by freezing the evolution of the KBO's eccentricity and apsidal line relative to the major axis of Planet Nine \citep{1985Icar...63..272W}. In particular, we compute the function 
\begin{align}
\Hres&= -\frac{\G\Msun}{2\,a} - n_9\bigg(\frac{\p}{\q}\bigg)\sqrt{\G\,\Msun\,a}\nonumber \\
&-\frac{\G\,m_9}{2\,\pi \,\q} \int_0^{2\pi \q} \bigg(\frac{1}{|\mathbf{r}-\mathbf{r}_9|} - \frac{\mathbf{r}\cdot \mathbf{r}_9}{|\mathbf{r}_9 |^3} \bigg)\bigg|_{\phires}\, d \lambda_9 ,
\label{Hres}
\end{align}
on a $(\phires-a)$ grid, setting the quantities $\Gamma$ and $\Delta\gamma$ to specific values. We note that unlike the doubly averaged Hamiltonian (\ref{Hpuresec}), here the averaging process is carried out only over $\lambda_9$, under the restriction of the resonant relationship dictated by equation (\ref{phi}). Moreover, the indirect part of the disturbing function must be retained in this case.

Of particular interest to the problem at hand are the resonant phase-space portraits of KBOs in apsidally anti-aligned configurations with respect to Planet Nine, with perihelion distances characteristic of typical scattered disk objects (i.e. $q\sim a_8$). Suitably, adopting $\Delta\gamma=\pi$ and a value for $\Gamma$ that corresponds to $q=35\,$AU at $a=(\ell/k)^{2/3}\,a_9$ for the frozen degree of freedom, we have computed the averaged Hamiltonian (\ref{Hres}) adopting parameters relevant to the 1:1, 3:2, 2:1, and 5:2 MMRs. The corresponding $\phires-a$ diagrams are presented in Figure (\ref{Fig:resonant}), on which we also plot $N-$body trajectories of particles shown in Figure (\ref{Fig:timeseries}) in red. It is clear that irrespective of the specific resonance argument, the topology of $\Ham_{\rm{res}}$ is keenly reminiscent of a pendulum-like structure, that has been cut into two separate domains by vertical lines. These lines depict \textit{collision curves} - i.e. values of $\phires$ for which the averaging process fails due to the inherent singularity. Importantly, this means that a trajectory residing within one domain cannot migrate to the other domain without compromising the phase-protection mechanism of mean-motion resonances. 

Within each of the phase-space portraits shown in Figure (\ref{Fig:resonant}), one of the two domains contains a $\infty$-shaped separatrix characterized by a hyperbolic equilibrium at its center, while the other domain possesses an elliptic fixed point at its core. Numerical integrations reported in the previous section have shown that resonant trapping typically occurs into the domain that does \textit{not} host the separatrix. That is, even though objects that exhibit stable libration of $\phires$ 180 degrees away from those shown in Figure (\ref{Fig:timeseries}) do exist, they are very rare and we will not consider their evolution in detail. We further note that strictly speaking, phase-space evolution depicted in Figure (\ref{Fig:resonant}) implies that the libration of $\phires$ does not represent a formal resonance because it is not enclosed by a separatrix \citep{2012A&A...546A..71D}. This point is however of little practical consequence, since the phase-protection facilitated by this pseudo-resonance patently represents a stabilizing mechanism for the simulated Kuiper Belt objects.

\subsection{Secular Dynamics Inside MMRs}\label{sect42}

Having characterized the evolution of the fast degree of freedom in the preceding subsection, we now consider secular dynamics facilitated by resonant interactions. Generally speaking, in order to compute a $e-\Delta\varpi$ phase-space diagram, we must specify the state of $a-\phires$ dynamics everywhere on the domain. To do so, we once again rely on the principle of adiabatic invariance.

Because the period of oscillation of $\Delta\varpi$ greatly exceeds that of $\phires$, we can define the adiabatic invariant \citep{1984PriMM..48..197N}
\begin{align}
\mathcal{J}= \oint \Lambda' \,d\phires' = -\frac{1}{\q} \oint \Lambda \,d\phires,
\label{adinv}
\end{align}
associated with motion in the $a-\phires$ plane. Physically, this quasi-integral corresponds to the phase-space area occupied by the trajectory, and is conserved to an excellent approximation, as long as the system does not encounter any criticality (i.e. hyperbolic fixed points, collision curves, etc; \citealt{Henrard1993}). As can be seen from Figure (\ref{Fig:timeseries}), bodies entrained in MMRs with P9 can have substantial libration amplitudes that are in essence determined by the state of the system at $t=0$. For definitiveness, here we ignore this complication and instead assume that the libration amplitude is null, meaning $\mathcal{J}=0$. From a computational point of view, the $\mathcal{J}=0$ assumption is simplifying, since rather than finding the correct phase-space trajectory in $a-\phires$ plane (specified by a given non-zero value of $\mathcal{J}$) at every combination of $e$ and $\Delta\varpi$, we instead suppose that the system is adiabatically confined to the resonant fixed point, and carry out the averaging process under the resonant equilibrium condition \citep{1993Icar..102..316M}.

Because the resonant angle of interest $\phires$ only contains the longitude of perihelion of Planet Nine and not that of the KBO (equation \ref{phi}), the resonant equilibrium in $a-\phires$ plane always resides on the $a_{\rm{eq}}=(\q/\p)^{2/3}\,a_9$ line. On the contrary, the equilibrium value of $\phires$ itself shifts away from $\phires=0\deg$ or $\phires=180\deg$, in concert with oscillations of $\Delta\varpi$. Correspondingly, in order to compute the equilibrium value of $\phires$ as a function of $e$ and $\Delta\varpi$, we evaluated the function $\mathcal{H}_{\rm{res}}$ (equation \ref{Hres}) along the $a=a_{\rm{eq}}$ axis at every grid point on the ($e,\Delta\varpi$) plane, and found its local maximum.

Within certain regions of the $e-\Delta\varpi$ diagram, the collision curves, shown as black vertical lines on Figure (\ref{Fig:resonant}), can approach one-another such as to shrink the domain within which the resonant trajectory resides. This process is demonstrated in Figure (\ref{Fig:varpimod}), which shows a series of resonant $a-\phires$ diagrams corresponding to the 3:2 MMR at various values of $\Delta\varpi$. Moreover, for specific values of $e$ and $\Delta\varpi$, the two collision curves can cross, engulfing the pseudo-resonant equilibrium point, which we assume the system occupies. In other words, the phase-space portrait in $a-\phires$ dictates a locus on the $e-\Delta\varpi$ plane that cannot be crossed without compromising the phase-protection mechanism inherent to mean motion commensurabilities\footnote{We note that the resonant-secular diagrams depicted in Figure (\ref{Fig:secres}) strictly correspond to resonant orbits that encircle the elliptic equilibrium points in Figure (\ref{Fig:resonant}). The comparatively less common orbits that reside in the domain occupied by the $\infty$-shaped separatrix, are thus subject to a quantitatively different secular evolution.}. Consequently, even prior to computing the secular dynamics explicitly, we can identify a restricted domain on $e-\Delta\varpi$ plane, bounded by the collision locus, that can be stably explored by an an orbit trapped at the center of a mean motion resonance.

Generally speaking, the admissible domain shrinks as the integer ($\p-\q$) increases. Accordingly, the full $e-\Delta\varpi$ plane is stable for the 1:1 MMR, but the stability region tightly confined about the $e\sim0.85$, $\Delta\varpi\sim\pi$ point for the 5:2 resonance. For clarity, here we only compute the secular diagram within this stability domain. 

Expressing $\lambda$ as a function of $\lambda_9$ through the pre-computed stationary value of $\phires$, we calculate the averaged Hamiltonian
\begin{align}
\Hresec = &-\frac{1}{4}\frac{\G\,\Msun}{a}\frac{1}{\big(1-e^2\big)^{3/2}} \sum_{j=5}^{8}\frac{m_j\,a_j^2}{\Msun\,a^2} \nonumber \\
&+\dot{\varpi}_9 \, \sqrt{\G\,\Msun a}\big(1-\sqrt{1-e^2} \big) \nonumber \\
&-\frac{\G\,m_9}{2\,\pi\, \q} \int_0^{2\pi \q} \bigg(\frac{1}{|\mathbf{r}-\mathbf{r}_9|} - \frac{\mathbf{r}\cdot \mathbf{r}_9}{|\mathbf{r}_9 |^3} \bigg)\bigg|_{\phires}  d \lambda_9
\label{Hsecres}
\end{align}
on the admissible domain. Figure (\ref{Fig:secres}) shows the contours of the numerically averaged function (\ref{Hsecres}) for the 1:1, 3:2, 2:1 and 5:2 MMRs (recall that corresponding resonant phase-space portraits are shown in Figure \ref{Fig:resonant}). In addition to semi-analytic level curves of $\Hresec$, depicted with black lines and the background color-scale of the Figure, we have also over-plotted the $e-\Delta\varpi$ evolutions of the four particles emphasized in Figure (\ref{Fig:timeseries}) as orange curves. Clearly, the agreement between semi-analytic results and $N$-body simulations is satisfactory, and any mild quantitative discrepancies can likely be attributed to our assumption of null resonant libration amplitude in the perturbative calculation. 

Taken together, these results yield the following insight into the dynamical evolution of distant Kuiper belt objects induced by Planet Nine. First and foremost, long-term dynamical stability is facilitated by capture into mean-motion resonances. Objects that are not (fortuniously) scattered into mean motion commensurabilities with Planet Nine initially, are removed from the system by way of close encounters. Meanwhile, due to the specific (``corotation") nature of the resonant multiplets that guide the resonant motion, the evolution of distant KBO eccentricities and longitudes of perihelion are dominated by secular dynamics that ensue inside resonances. In turn, this separation between the degrees of freedom qualitatively explains why purely secular phase-space portraits shown in Figure (\ref{Fig:secular}) approximately match the results of large-scale numerical simulations. 

\subsection{Critical Semi-Major Axis}

In light of the analysis presented above, it is evident that even though the purely secular treatment of dynamics outlined in section (\ref{sect2}) does not formally account for the full dynamical evolution observed within $N$-body calculations, it does provide a satisfactory approximation for $e-\Delta\varpi$ evolution \citep{Beust2016}. Accordingly, we now take advantage of this simplified framework to explore the dependence of the critical semi-major axis, $a_{\rm{crit}}$ (beyond which apsidal confinement ensues) on the parameters of Planet Nine. 

As a proxy for $a_{\rm{crit}}$, we adopt the minimum value of $a$ at which the anti-aligned secular equilibrium exists on the secular $e-\Delta\varpi$ diagram\footnote{For our nominal parameters (i.e. $a_9=700\,$AU, $e_9=0.6$, $m_9=10\,m_{\oplus}$), this proxy yields $a_{\rm{crit}}\approx200\,$AU, in good agreement with simulations.}. Practically, we calculate this quantity by computing $\Hsec$ as a function of $e$ along the $\Delta\varpi=\pi$ line, gradually increasing $a$ from $50\,$AU, and noting the first instance where a local maximum appears between $e=0$ and $e=1$. Correspondingly, this calculation is carried out on a ($a_9,e_9$) grid for a given value of $m_9$.

\begin{figure}[t]
\centering
\includegraphics[width=1\columnwidth]{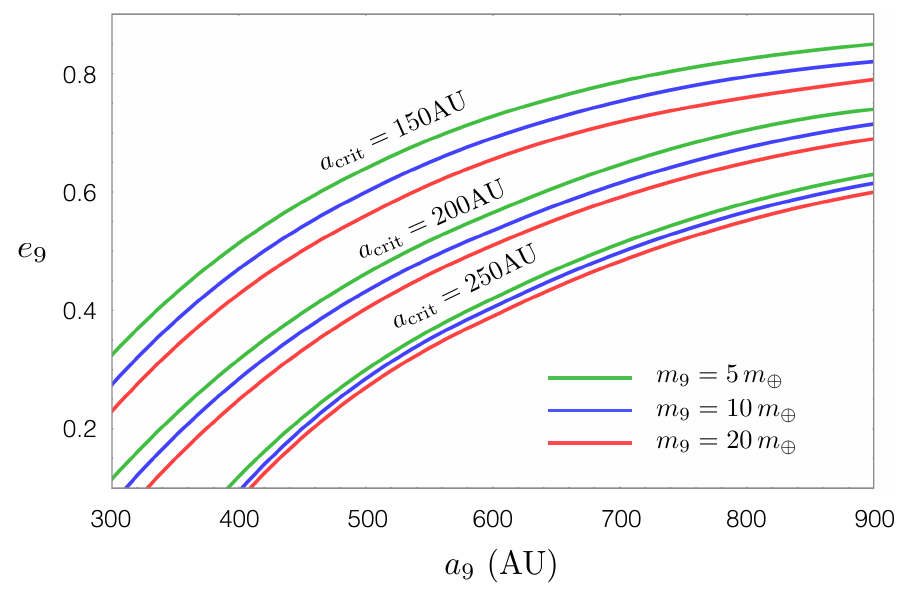}
\caption{Critical semi-major axis corresponding to a transition from randomized to apsidally clustered regime of KBO dynamics. In light of the observational ambiguity related to the specific value of $a_{\rm{crit}}$, here we show contours corresponding to $a_{\rm{crit}}=100$, $150$, and $250\,$AU as functions of $a_9$, and $e_9$, for $m_9=5$ (green), $10$ (blue), and $20\,m_{\oplus}$ (red). The calculation is carried out in the secular approximation, and assumes that Planet Nine's inclination is not sufficiently large to alter the $e-\Delta\varpi$ dynamics appreciably.}
\label{Fig:acrit}
\end{figure}

We have computed $a_{\rm{crit}}$ as a function of $a_9$ and $e_9$ for $m_9=5\,m_{\oplus}$, $m_9=10\,m_{\oplus}$, and $m_9=20\,m_{\oplus}$. Figure (\ref{Fig:acrit}) depicts curves corresponding to $a_{\rm{crit}}=150\,$AU, $a_{\rm{crit}}=200\,$AU, and $a_{\rm{crit}}=250\,$AU for these choices of $m_9$. As can be clearly seen in this figure, $a_{\rm{crit}}$ exhibits a rather mild dependence of $m_9$, and follows a shallow relationship between $e_9$ and $a_9$.
 
It is important to note that observationally, the specific value of $a_{\rm{crit}}$ is not unequivocally determined. It is indeed possible to demonstrate that the clustering of orbits with $a\gtrsim250\,$AU in their respective \textit{longitudes} of perihelion ($\varpi$) is statistically significant \citep{2017AJ....154...65B}. However, given that the confinement in the \textit{argument} of perihelion ($\omega$) may persist down to $a\sim150\,$AU \citep{TS14}, it is possible that anti-aligned dynamics in fact ensues beyond $a>150\,$AU, and the observed $\varpi$ distribution in the $150<a<250\,$AU range is contaminated by metastable objects residing within the apsidally \textit{aligned} regions of the $e-\Delta\varpi$ diagram. While perfectly plausible, this scenario is not as strongly supported in a raw statistical sense by the current dataset \citep{2017AJ....154...65B,Shankman}, leaving some ambiguity as to where in the $150<a<250\,$AU interval the true value of $a_{\rm{crit}}$ resides. 

Whatever the exact value of $a_{\rm{crit}}$ may be, it is worth noting that P9 parameters within $N$-body simulations that were found to match the observational data with comparatively high probability by \citet{BroBat} all lie between the $a_{\rm{crit}}=150\,$AU and $a_{\rm{crit}}=250\,$AU contours in Figure (\ref{Fig:acrit}). In other words, these contours delineate a region of orbital element space that yields simulation results that compare well with the real solar system. At the same time, the semi-analytic calculations presented herein do not suffer from limitations in resolution (i.e. particle count) inherent to numerical experiments, and therefore likely inform a broader range of acceptable P9 parameters than was reported in \citet{BroBat}. Exploring this parameter range with an expanded suite of high-resolution $(N\gg5000)$ $N$-body simulations constitutes an appealing avenue for future development of the Planet Nine hypothesis.

\section{Spatial Dynamics}\label{sect5}

Up until this point, we have considered the orbital evolution induced upon the Kuiper belt by Planet Nine, under the strict assumption of coplanarity. As already discussed above, this assumption leads to somewhat idealized behavior, and fails to capture three important aspects of the dynamics that emerge within the framework of full-fledged $N$-body simulations. First, rather than exhibiting stable resonance capture and remaining locked to a particular commensurability for the duration of the solar system's lifetime, real Kuiper belt objects experience chaotic semi-major axis evolution, and therefore explore a wide range of orbital period ratios with Planet Nine (this is evident, for example, in Figure 11 of \citealt{ML2017}). Second, in addition to confinement in the longitude of perihelion, real long-period KBOs exhibit a clustering in the longitude of ascending node as well, which jointly leads a clustering in the argument of perihelion, $\omega=\varpi-\Omega$. Third, $N$-body simulations of \citet{BB16} show that Kuiper belt objects initially residing close to P9's orbital plane can occasionally undergo high-amplitude oscillations of the inclination, leading to generation of retrograde KBOs \citep{BB16b}. Accordingly, the goal of this section is to characterize these features qualitatively.

\subsection{The Anti-Aligned Population}

To carry out the exploration of non-planar dynamics within the framework of our simplified model, we repeated the numerical experiment with $a_9=700\,$AU described in section \ref{sect3}, allowing for a small inclination dispersion among the particles. In particular, initial particle inclinations were drawn from a half-normal distribution with a standard deviation of $\sigma_i=5\deg$, while longitudes of ascending node $\Omega$ were assumed to be uniformly distributed between $0$ and $360\deg$. As before, the simulations are performed in a frame coplanar with the orbit of Planet Nine (meaning that $i_9=0$), and the Keplerian motion of the four inner giants is averaged out. Meanwhile, the ecliptic plane is envisioned to be inclined with respect to the orbit of P9 by a small amount (e.g. $10-20\deg$), although the corresponding reduction in the effective magnitude of the $J_2$ moment (equation \ref{Jeff}) is ignored. Remarkably, this simple modification to the physical setup of the problem is sufficient to approximately recover the full breadth of the dynamical behavior observed within more detailed numerical experiments. 

\begin{figure}[t]
\centering
\includegraphics[width=1\columnwidth]{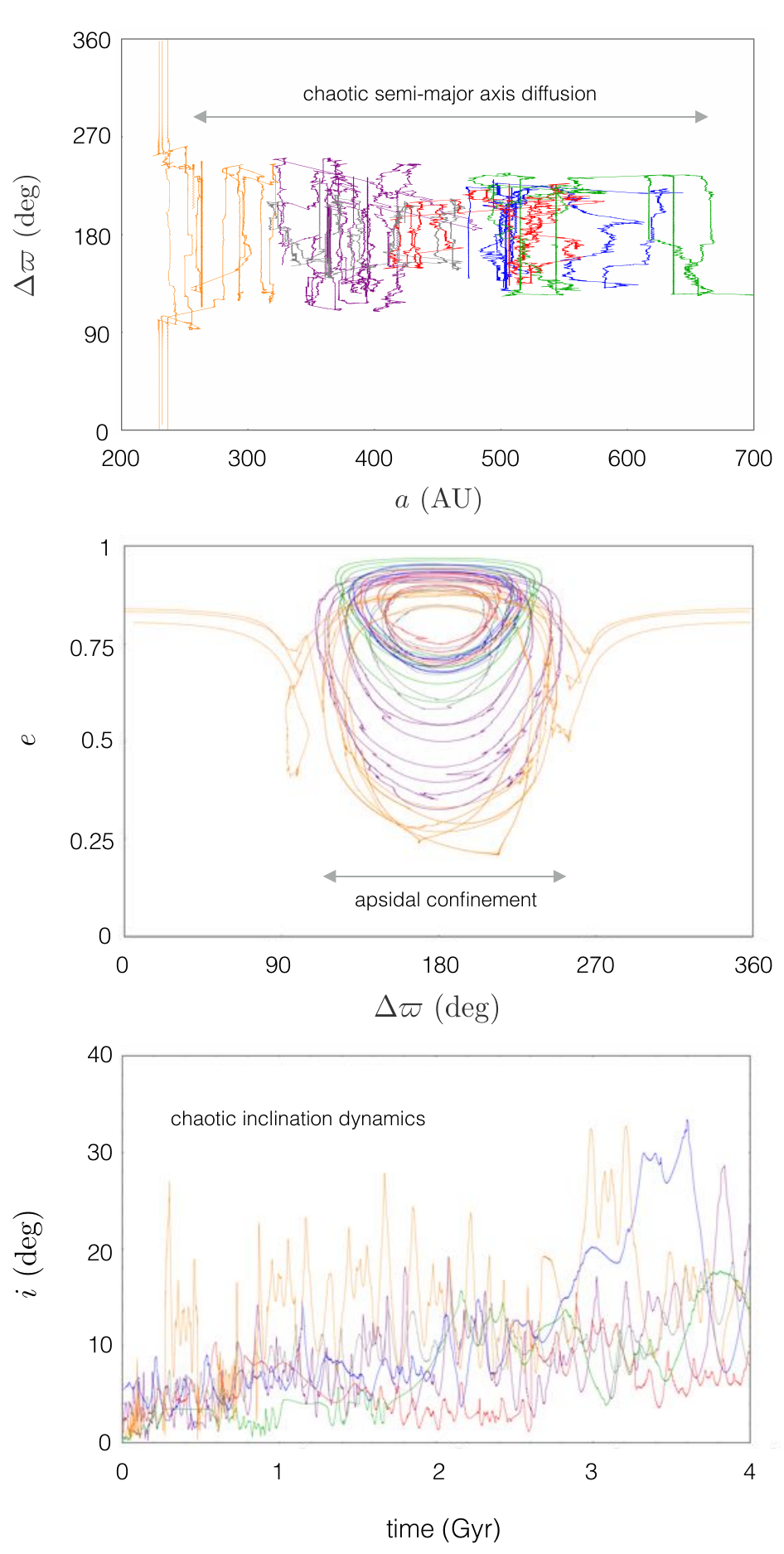}
\caption{Dynamical evolution exhibited by apsidally confined objects that do not experience large-amplitude inclination oscillations. The top panel shows the chaotic footprint outlined by six uniquely colored particles in $\Delta\varpi-a$ space, and elucidates the fact that a small inclination dispersion renders the semi-major axis evolution of distant Kuiper belt objects stochastic. Each vertical segment, however, traces out long-term trapping of particles in mean motion resonances. The middle panel displays $e-\Delta\varpi$ dynamics, demonstrating that secular evolution experienced by the particles retains its apsidally clustered character, despite chaotic variations of the semi-major axes, except when $a$ approaches $a_{\rm{crit}}$. The bottom panel depicts chaotic evolution of orbital inclinations. In a reference frame that coincides with Planet Nine's orbit, the angular momentum vectors of distant KBOs chaotically rotate around the orbit normal. Viewed from the ecliptic plane, however, circulation of distant object around Planet Nine's mildly inclined plane will yield an apparent clustering of the longitudes of ascending node.}
\label{Fig:clustered}
\end{figure}

Figure (\ref{Fig:clustered}) shows the orbital evolution of six (uniquely colored) representative objects, whose orbits remain roughly confined to the orbital plane of Planet Nine. The top panel depicts the longitude of perihelion (relative to the apsidal line of Planet Nine) as a function of semi-major axis. Clearly, allowing the KBOs to possess even a small inclination breaks the immutability of mean motion resonances seen in the strictly coplanar model, and renders the evolution of the semi-major axes chaotic. At the same time, the clustering of the orbits in longitude of perihelion persists in spite of irregular semi-major axis evolution, and indeed, the secular portrait of orbital eccentricity depicted in the middle panel of the Figure agrees well with the resonant-secular $e-\Delta\varpi$ diagrams shown in Figure (\ref{Fig:secres}).

We note that although the introduction of a finite inclination dispersion is sufficient to drive chaotic mixing in the semi-major axis, its efficiency is underestimated in our simplified model, since stochastic evolution is driven solely by P9 resonances. In reality, scattering facilitated by Neptune significantly enhances the rate of semi-major axis diffusion (particularly during the low-perihelion phases of the secular cycle), and modifies the observed behavior on a quantitative level. As a consequence, the results obtained in our semi-averaged simulations depict a somewhat idealized realization of distant Kuiper belt evolution.

The apsidally clustered resonance-hopping behavior observed in Figure (\ref{Fig:clustered}) can be qualitatively understood within the framework of the semi-analytical theory described above. The introduction of a new ($i-\Omega$) degree of freedom into the dynamics implies that rather than being forced by a single secular harmonic, the modulation of the $a-\phires$ phase-space portrait is now driven by two separate angles. This modulation transforms the collision loci depicted in Figure (\ref{Fig:secres}) into fuzzy chaotic bands because any given point ($e, \Delta\varpi$) near the original collision locus may or may not lead to a collision, depending on the values of $i$ and $\Omega$. To this end, notice that in Figure (\ref{Fig:secres}), the curves showing the secular evolution are essentially tangent to the collision loci, which means that in a strictly planar physical setup, secular evolution almost never leads to close encounters between the particles and Planet Nine (i.e. most of the plotted trajectories don't intersect the collision loci). On the other hand, if the collision locus becomes a chaotic band, many more secular trajectories can infiltrate this irregular region, and with appropriate values of $i$ and $\Omega$, close encounters can ensue. When this happens, the particle enters a stochastic dynamical regime and hops in semi major axis, until it locks into a new resonance and the secular dynamics drives it away from the collision locus of the new resonance. Importantly, this type of behavior is observed in numerical integrations of real objects under the influence of Planet Nine \citep{ML2017,Becker2017}.

When an object gets fortuitously trapped in some resonance (by entering it through the chaotic layer that surrounds the collision curves), its orbital evolution can be temporarily stabilized by the secular evolution in $e$ and $\Delta\varpi$. In other words, test-particles have the tendency to exhibit prolonged periods of resonance locking (on timescales similar to the secular libration period) before breaking out, and jumping to another commensurability. In fact, looking again at Figure (\ref{Fig:secres}) and imagining a chaotic band near each depicted collision locus, it is evident that the ($e,\Delta\varpi$) evolution can drive a body away from the band at the peaks of the eccentricity cycle of the 3:2, 5:3 and 2:1 resonances (and also at the bottom of the eccentricity cycle in the 2:1 resonance). During this secular phase, close encounters between the particles and Planet Nine are no longer possible. However, as the eccentricity-perihelion cycle unfolds, the particle must eventually plunge back into the chaotic band. The object can thus experience new close encounters with Planet Nine, and hop to another resonance where this process repeats. In reality, this sequence of events is further complicated by the fact that at the peak of the eccentricity cycle, objects are brought to an orbital state where $q\sim a_8$ and suffer enhanced semi-major axis diffusion due to scattering off of Neptune (or equivalently, overlap with Neptune's exterior mean motion resonances; \citealt{2008ssbn.book..259G}). As a result, it is reasonable to assert that only dynamically ``detached" objects that are not actively scattering off of Neptune, are presently entrained in mean motion resonances with Planet Nine.

The dynamical evolution of the orbital inclination observed in the bottom panel of Figure (\ref{Fig:clustered}) is a consequence of the chaotic rotation of the angular momentum vectors around Planet Nine's orbit normal. That is, their longitudes of ascending node relative to the orbital plane of Planet Nine are in circulation. However, if the inclination of P9 is sufficiently large relative to the ecliptic, then particles that are less inclined with respect to Planet Nine's plane than the inclination of P9 itself, will appear to have a librating node relative to the ecliptic. In other words, viewed from a coordinate system that coincides with the ecliptic plane, the orbits of these particles execute a libration around the forced $i - \Delta\Omega$ equilibrium forced by P9's inclination.

Taken together, our calculations suggest that the clustering of the longitudes of ascending node first noted in \citet{BB16} is nothing but a trivial consequence of the bending of the Laplace plane away from the solar system's mean plane by Planet Nine. Moreover, the apparent libration of the ascending node $\Omega$, together with the true libration of longitude of perihelion $\varpi$, produces the apparent libration of the argument of perihelion $\omega$, as observed in the real data \citep{TS14}. This implies that the orbital inclination of Planet Nine must simultaneously be sufficiently large for apparent nodal clustering to ensue (e.g. $i_9\gtrsim10-20\deg$), but not be so large as to disrupt the stable confinement of the longitudes of perihelion ($i_9\lesssim40\deg$; \citealt{BroBat,2017arXiv170701379S}).

\subsection{The Highly-Inclined Population}\label{sect52}

Perhaps the most remarkable consequence of P9-driven dynamics is exemplified by the induction of large-amplitude oscillations in the orbital inclinations of distant KBOs. Not only is this mode of orbital evolution a unique prediction of the Planet Nine hypothesis \citep{BB16}, real objects presently entrained in this pattern of perturbation now comprise a firmly established part of the observational dataset \citep{Gomes2015}. Accordingly, this regime of P9-induced dynamics constitutes one of the strongest lines of evidence for the existence of Planet Nine, as no other dynamical model can reasonably account for the origin of the observed highly inclined TNO population. Let us now examine these extreme orbital excursions within the framework of our simplified numerical model. 

\begin{figure}[t]
\centering
\includegraphics[width=1\columnwidth]{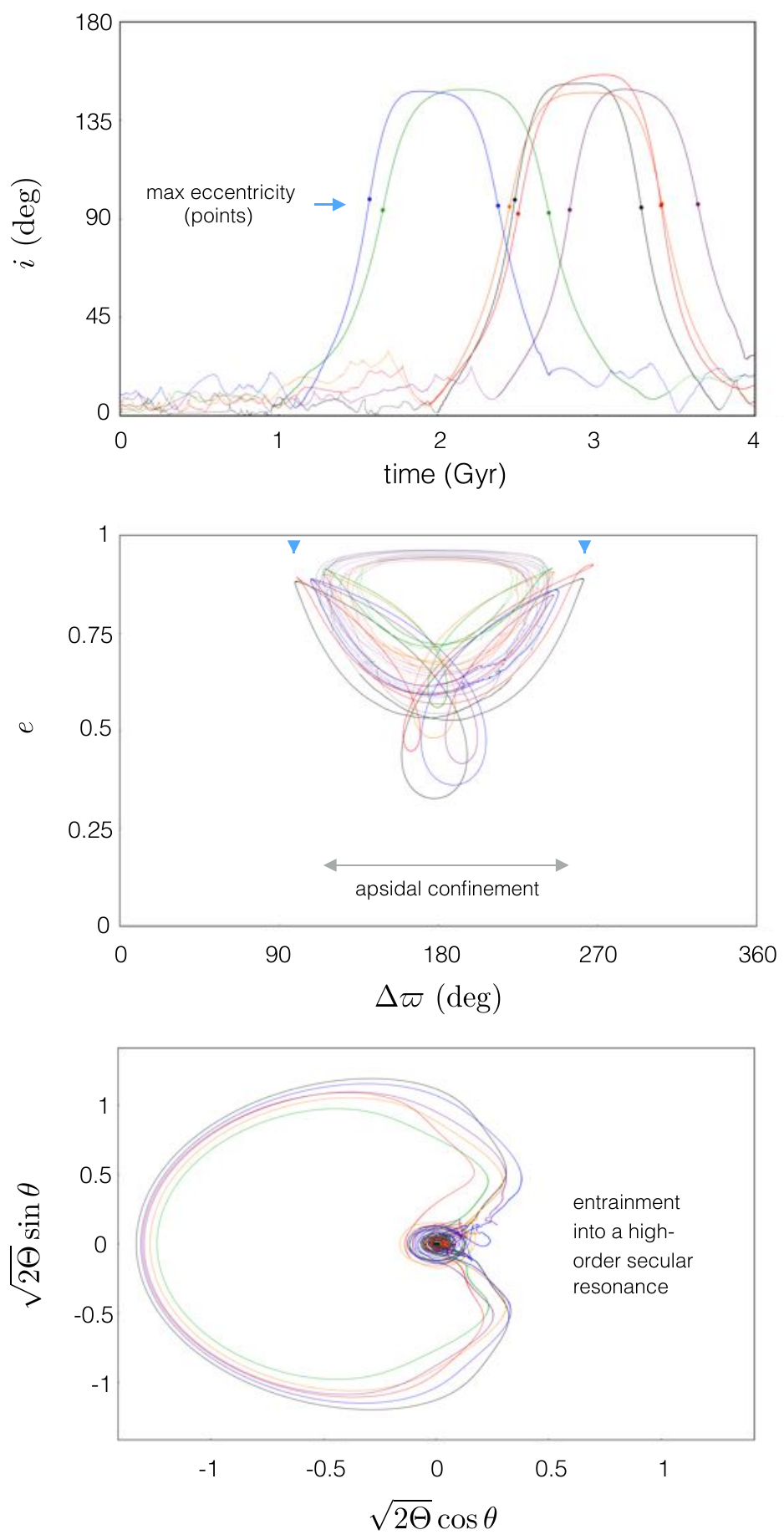}
\caption{Dynamical evolution exhibited by simulated particles that experience orbital flips. The top panel shows the inclination time series of six uniquely colored test particles that exhibit large-amplitude orbital excursions and temporarily achieve retrograde orbits. As the large-scale variations of inclination ensue, $e-\Delta\varpi$ projection of the dynamics acquires a distinct three-lobed shape, which is characterized by eccentricity maxima that are approximately $70\deg$ away from $\Delta\varpi=180\deg$, and are achieved when $i\approx90\deg$. In contrast with Kozai-Lidov dynamics, the depicted evolution is characterized by simultaneous libration of the critical angles $\theta$ and $\Delta\varpi$ (as shown on the bottom panel), and constitutes an exceedingly strong form of secular coupling.}
\label{Fig:inclined}
\end{figure}

The top panel of Figure (\ref{Fig:inclined}) shows the inclination time series of six simulated particles with initial semi-major axes between $500\,$AU and $600\,$AU that do not remain bound to Planet Nine's orbital plane for the entire duration of the integration. As can be readily seen in this Figure, during the latter half of the solar system's lifetime, each object abruptly enters a phase of extreme orbital variation, and upon experiencing a single large-scale oscillation of the inclination rejoins the low-$i$ population of apsidally anti-aligned bodies. While representative, we note that orbital excursions of this sort are not always limited to a single cycle - some objects within the simulation suite experience a multitude of sequential oscillations. Moreover, the onset of high-$i$ excursions is not limited to a small subset of particles - in our simulation, 38\% of all stable objects experience at least one such excursion.

An intriguing feature of the depicted evolution is that the pattern of $e-\Delta\varpi$ dynamics changes drastically when a particle enters the highly inclined regime. As shown on the middle panel of Figure (\ref{Fig:inclined}), rather than encircling an elliptic equilibrium located at $\Delta \varpi=180\deg$ as in Figure (\ref{Fig:clustered}), the $e-\Delta\varpi$ projection of the phase-space portrait acquires a three-lobed shape with eccentricity maxima located $\sim\pm70\deg$ away from perfect apsidal anti-alignment with Planet Nine. The corresponding instants where the eccentricities are maximized (and perihelion distance is minimized) are shown with points on the top panel of Figure (\ref{Fig:inclined}), and lie almost exactly at $i\approx90\deg$.

Taken together, the top and middle panels of Figure (\ref{Fig:inclined}) show that the model predicts the highly inclined population to be most readily observable in a state that is approximately perpendicular to the ecliptic, and is slightly sub-orthogonal in apsidal orientation with respect to the anti-aligned cluster of distant orbits. Note however, that during phases of lower eccentricities (and higher perihelion distance), this population remains relatively well localized in the longitude of perihelion\footnote{Recall that the longitude of perihelion is a dog-leg angle, and for highly inclined orbits does not generally represent a good proxy for the azimuthal angle of the Runge-Lenz vector.} around $\Delta\varpi=180\deg$. This means that Kuiper belt orbits that lie beyond the current observational frontier exhibit an even more complex dynamical structure  than those comprising the known long-period dataset.

While it is tempting to attribute the mode of orbital evolution shown in Figure (\ref{Fig:inclined}) to the oft-cited Kozai-Lidov (KL) resonance \citep{1962P&SS....9..719L,1962AJ.....67..591K}, it is crucial to understand that the flavor of secular dynamics executed by the simulated particles is keenly distinct. Specifically, in contrast with conventional KL evolution (where orbital inclination is traded for eccentricity such that the particle's eccentricity is minimized when orbits become orthogonal), the simulated particles reach their peak eccentricities near $i\approx90\deg$. Moreover, instead of being constrained by the conservation of the $\hat{z}-$component of the specific angular momentum vector $h=\sqrt{1-e^2}\cos(i)$ \citep{1999CeMDA..75..125K}, the orbital excursions observed in our simulation are accompanied by large-amplitude variations of this quantity (ranging from $\sim1$ to -1). This further implies that the KL resonance does not represent the primary driver of the depicted evolution.

Because the orbit of Planet Nine is assumed to have $i_9=0$ in our calculations, the only angles that appear in the secular Hamiltonian are $\Delta\varpi$, $\omega$, and their various linear combinations. We have already argued above that the KL mechanism is not responsible for the observed evolution, so libration of $\omega$ alone cannot facilitate the observed dynamics\footnote{Recall that the critical angle associated with the KL resonance is $2\omega$ \citep{1999CeMDA..75..125K}.}. Simultaneously, $\Delta\varpi$ cannot force oscillations in inclination, meaning that the secular resonance at play must contain both $\Delta\varpi$ and $\omega$. Correspondingly, we propose that large-scale orbital variations depicted in Figure (\ref{Fig:inclined}) are driven by libration of the secular angle 
\begin{align}
\theta=\Delta\varpi-2\omega=2\Omega-\varpi-\varpi_9.
\label{theta}
\end{align}
Incidentally, this angle arises at the octopole order of expansion when the Hamiltonian is expressed as series in semi-major axis ratios \citep{2010MNRAS.407.1048M}.

Adopting $\Delta\gamma=-\Delta\varpi$ and $\theta$ as the secular angles of the test particles, we identify the quantity
\begin{align}
\Theta=\frac{\sqrt{1-e^2}}{2}\left(1-\cos i \right)
\label{Theta}
\end{align}
as the action conjugate to $\theta$. In the bottom panel of Figure (\ref{Fig:inclined}), we show the evolution of canonical cartesian coordinates related to $(\theta,\Theta)$ action-angle variables. From this figure, it is evident that during the high-inclination phase of orbital evolution, $\theta$ executes a bounded oscillation and the secular trajectory traces out the shape of a typical resonant separatrix \citep{1983CeMec..30..197H}. Thus, the dynamics shown in Figure (\ref{Fig:inclined}) is characterized by the simultaneous libration of the relative longitude of perihelion $\Delta\varpi$, as well as the angle $\theta$ (which jointly leads to the libration of the longitude of ascending node) representing an exceptionally strong form of secular coupling.

Due to the lack of separation of timescales on which the two secular degrees of freedom operate, we cannot study the depicted large-amplitude variations of inclination using the same flavor of semi-analytic perturbation theory as that outlined in section \ref{sect4}. Nevertheless, we note that the onset of these large-scale oscillations almost always corresponds to the eccentricity minimum in the $e-\Delta\varpi$ cycle that precedes the oscillation. Qualitatively, this implies that resonant-secular $e-\Delta\varpi$ dynamics modulates the proximity parameter of the secular $\theta-\Theta$ resonance, such that when $e$ approaches a critical value, a dynamical gateway towards capture of low-inclination objects into the secular resonance characterized by libration of $\theta$ temporarily opens. Accordingly, the coercion of low-$i$ objects onto trajectories that experience large-amplitude oscillations of inclination is a fundamentally stochastic process. In turn, this means that some of the currently observed members of the apsidally clustered population can in principle join the highly inclined population in the future, and visa-verse. 

\section{Comparison with Observations}\label{sectobs}

Given that our spatial model reproduces the key features of more detailed numerical simulations, it warrants a rudimentary comparison with the observational data in light of the semi-analytical insight into the governing dynamics developed above. Because we are neither resolving the Keplerian motion of the canonical giant planets, nor their inclination with respect to Planet Nine, here we will not consider the clustering of the longitudes of ascending node at low $i$. Instead, we will focus exclusively on the confinement of the longitude of perihelion as well as the behavior of the highly inclined long-period objects. 

As shown in Figure (\ref{Fig:orbits}), there are currently ten known objects with $a>250\,$AU (shown in purple) that comprise the primary $\varpi$ cluster. The observational dataset also shows the existence of two objects that are diametrically opposed to the mean orientation of this cluster (shown in green), as well as a single outlier, 2015\,GT$_{50}$ (shown in gray), that does not fall within either the apsidally aligned or anti-aligned sub-populations of objects. In order to meaningfully compare the expectations of the model with the data, we extended our integrations such that the initial semi-major axis distribution of the particle disk stretches out to $850\,$AU.

\begin{figure}[t]
\centering
\includegraphics[width=1\columnwidth]{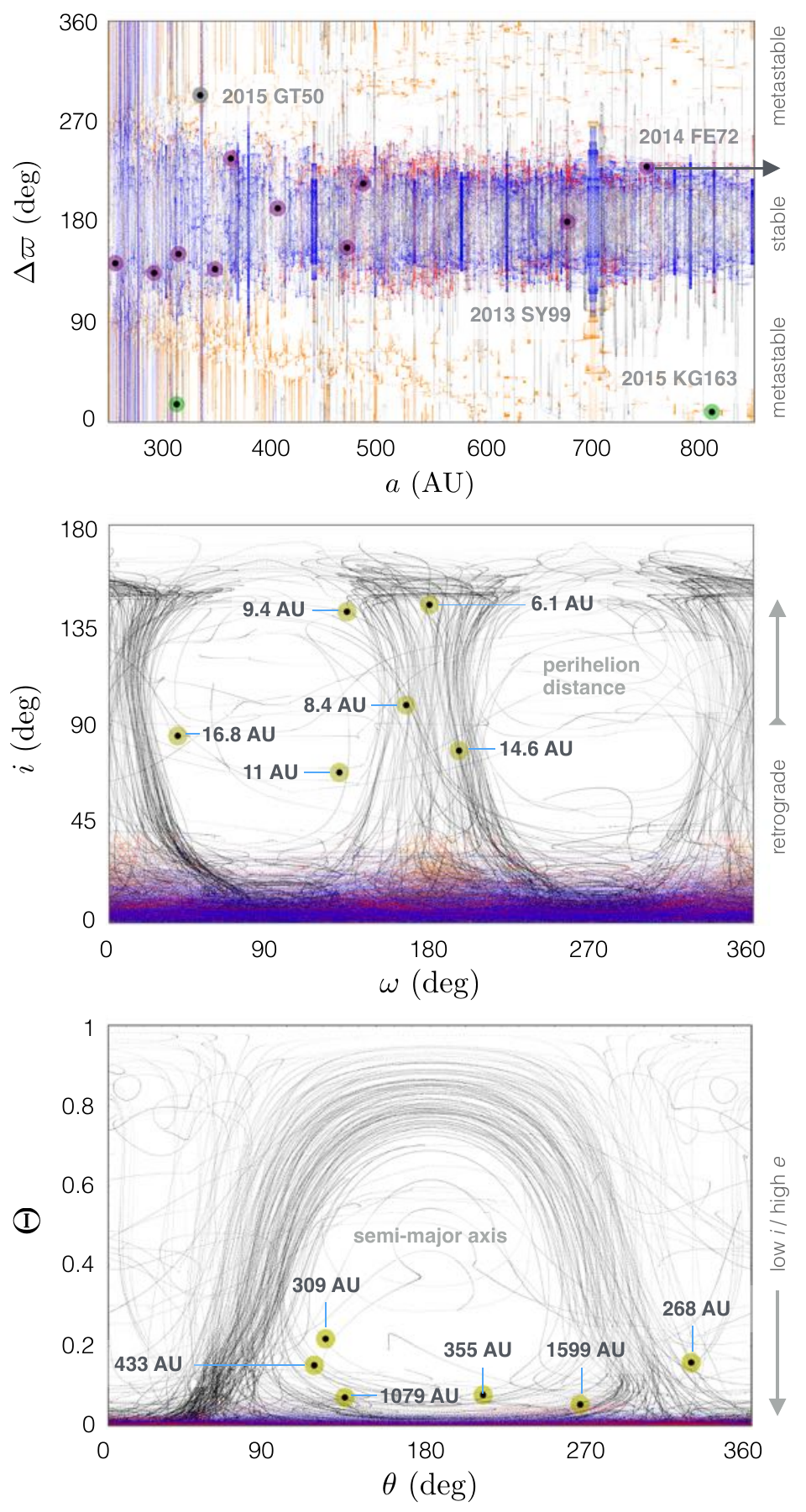}
\caption{Comparison between the results of our $N$-body simulation with $a_9=700\,$AU, $e_9=0.6$, $m_9=10\,m_{\oplus}$ and the observational data. The top panel shows the chaotic $\Delta\varpi-a$ footprint outlined by dynamically long-lived and metastable particles within the simulation. Evolution of long-term stable low-$i$ objects is displayed with blue dots when they satisfy our crude observability criteria, and with gray dots when they attain $q>100\,$AU or $i>40\deg$. Similarly, objects that do experience large-scale inclination cycles are shown with red dots when visible and with gray dots otherwise. Meanwhile, the orbital footprint of metastable objects with dynamical lifetimes between $100$ and $500\,$Myr are shown with orange dots. The current observational census of distant Kuiper belt objects is over-plotted on the panel, and is color-coded in the same way as in Figure (\ref{Fig:orbits}). Note that 2014\,FE$_{72}$ has a semi-major axis of $1923\,$AU, and is shown on the figure for completeness. The middle panel depicts the orbital inclination of simulated particles as a function of their argument of perihelion, while the bottom panel projects the same trajectories onto a plane defined by the action-angle coordinates $\Theta$ and $\theta$. Six currently known long-period centaurs are over-plotted on the figure as yellow points, and their perihelion distances and semi-major axes are labeled.}
\label{Fig:datasim}
\end{figure}

The orbital footprint of all simulated long-term stable and metastable particles is shown in Figure (\ref{Fig:datasim}), where the top panel depicts longitude of perihelion as a function of the semi-major axis (as in the top panel of Figure \ref{Fig:clustered}). Meanwhile, the middle panel shows the orbital inclination as a function of the argument of perihelion and the bottom panel elucidates the the action $\Theta$ as a function of its conjugate angle, $\theta$. As a crude proxy for observability, we adopt simple cuts of the numerical output at $q\leqslant100\,$AU and $i\leqslant40\deg$. Points corresponding to bodies with dynamical lifetimes in excess of $4\,$Gyr that do not simultaneously satisfy these criteria are shown in gray, while those that do are shown in blue or red, depending on their inclination evolution. In particular, objects that remain confined to the plane of Planet Nine throughout the integration are shown in blue. On the other hand, bodies that experience large-amplitude inclination oscillations at any point in their evolution are depicted in red. 

In agreement with the results of \citet{BB16}, the top panel of Figure (\ref{Fig:datasim}) shows the emergence of a well-defined cluster of apsidally anti-aligned orbits that are contaminated by trajectories that circulate in the longitude of perihelion. One striking example of such a circulating trajectory is shown as a vertical blue line with a semi-major axis of $a\approx335\,$AU. As discussed in sections \ref{sect3} and \ref{sect4}, the dynamics of apsidally confined trajectories are driven by the resonant harmonic $\phires$ (see equation \ref{phi}), while the comparatively less frequent apsidally circulating trajectories tend to reside within resonant multiplets characterized by low-amplitude libration of other resonant angles that contain the particle's longitude of perihelion, $\varpi$. The aforementioned observational data are over-plotted on this panel, and are color-coded in the same way as in Figure (\ref{Fig:orbits}). 

As expected, all observed KBOs that comprise the primary cluster (shown in purple) are seamlessly explained by the simulation results. More remarkably, however, the outlier within the data (shown in gray) is also naturally reproduced by the model, as an object that belongs to the class of observable stable particles that exhibit apsidal circulation, i.e. those entrained in resonances characterized by angles that contain $\varpi$ (see appendix for an analysis). In the present example, this agreement stems from the fact the observed object is very close to the 3:1 resonance with Planet Nine. While this correspondence may be purely accidental, it emphasizes that the mere existence of a small number of apsidally unconfined objects that do follow the overall pattern exhibited by the data, does not constitute strong evidence against the Planet Nine hypothesis.

Although a similar narrative could in principle be invoked for the two objects that are the apsidally aligned with Planet Nine (shown in green), it can be more reasonably speculated that these bodies are in fact subject to purely secular interactions with P9. Recall from section \ref{sect2} (Figure \ref{Fig:secular}) that apsidally aligned objects residing in the secular domain are protected from close encounters by the geometric co-linearity of the orbits. Correspondingly, objects that never attain low perihelion distances at the top of their eccentricity cycle and are therefore close to the secular equilibrium point, exhibit long-term stable apsidal libration about $\Delta\varpi=0$. Conversely apsidally aligned objects with low perihelion distances are metastable, as they only begin to scatter off of Planet Nine once they precess onto their respective tangential collision curves, yielding dynamical lifetimes that are on the order of (a fraction of) the precession timescale - i.e., $\sim$ few$\,\times\,100\,$Myr. 

Because in our numerical experiments we restricted the initial perihelion range of all particles to $q\in(30,36)\,$AU, our simulations do not produce any long-term stable apsidally aligned orbits. Instead, secular dynamics observed within our $N$-body simulations are dominated by the metastable particles that originate at high eccentricity and eventually precess towards the tangential configuration, where they are scattered out. To demonstrate the apsidal behavior of this sub-population of orbits, objects with dynamical lifetimes between $100$ and $500\,$Myr are shown as orange dots in Figure (\ref{Fig:datasim}). These particles clearly cluster around $\Delta\varpi=0$, and provide an excellent match to the aligned (green) data points shown in the Figure. 

Naturally, this explanation would not be sensible if the entire distant Kuiper belt had been generated $\sim4\,$Gyr ago and never replenished since then. This is however not the case in the real solar system: just as the highly-inclined long-period KBOs are routinely scattered inwards to create the population of retrograde bodies with $a<100$\,AU \citep{BB16b}, scattered disk objects with $a<250\,$AU are continuously scattered outward by Neptune, resupplying the distant trans-Neptunian region with metastable KBOs \citep{2008ssbn.book..259G}. Thus, our model points to the possibility that 2013\,FT$_{28}$ and 2015\,KG$_{163}$ are relative newcomers to the distant Kuiper belt, and will eventually be destabilized by short-periodic interactions with Planet Nine.

There exists yet another interpretation of the apsidally aligned data points as well, which is not well represented by Figure (\ref{Fig:datasim}) due to our choice of low-$q$ initial conditions. Particularly, rather than belonging to the aforementioned metastable sub-population of bodies that experience pure secular evolution, 2013\,FT$_{28}$ and 2015\,KG$_{163}$ could be entrained in the stable secular libration island around $\Delta\varpi=0$, and are presently observed near the peak of their respective eccentricity cycles. In order to unequivocally distinguish between these two interpretations, we would need to know the exact orbital parameters of Planet Nine. However, we simultaneously note that within the context of the long-term stable interpretation, some additional mechanism other than scattering off of Neptune (such as say, interactions with the birth cluster; \citealt{2004AJ....128.2564M,2010ARA&A..48...47A}) would likely be required to initially raise the perihelion distances of these objects and lock them into the apsidally aligned secular libration island. This is because bodies scattered to distant elliptic orbits by the traditional giant planets would necessarily have low perihelia (like the initial conditions of our simulation) and therefore could not reside within the stable libration island.

The yellow data points shown on the middle and bottom panels of Figure (\ref{Fig:datasim}) represent the population of distant ($a>250\,$AU), highly inclined ($i>40\deg$) objects with $q<30\,$AU discussed in section \ref{sect52}. Although these objects conform to the dynamical streamlines traced out by the simulated particles relatively well, it is important to keep in mind that (by virtue of having $q<30\,$AU) these bodies have eccentricities that are much closer to unity than any of the particles in our simulations. This means that the observed objects are drawn from the extreme end of the broader high-$i$ population, and have likely had their orbits somewhat perturbed by the canonical giant planets. As a result, we expect that the agreement between theory and observations will be even better for (yet-undiscovered) highly inclined long-period KBOs with $q>30\,$AU. Certainly, continued observational monitoring of the distant Kuiper belt outside of the ecliptic plane constitutes a viable avenue towards further characterization of long-term dynamical evolution induced by Planet Nine.

As a final point, it is instructive to remark on the extent to which the calculations described above are in agreement with more detailed simulations that fully resolve the orbital motion of the inner giants. In particular, the top panel of Figure (\ref{Fig:datasim}) can be readily compared with Figures (5) and (8) of \citet{BB16}, which depict simulations with very similar initial conditions to that considered herein. As expected, upon comparison of these numerical experiments we find that the general survival rate of particles is somewhat lower in simulations that model Neptune directly. Specifically, in our semi-averaged calculations, 19\% of particles initialized between 250\,AU and 550\,AU remain stable over 4Gyr, while only 7\% of objects in this initial semi-major axis and perihelion range survive the full integration span in a simulation where Neptune is modeled directly. This number increases slightly to 10\% when Planet Nine is endowed with a $i_9=30\deg$ inclination with respect to the inner solar system.

Another notable difference between direct and semi-averaged calculations lies in that simulated objects that circulate in perihelion and match our observability criteria are somewhat less prevalent in the simulations that include Neptune's short-periodic perturbations. This is likely because apsidally circulating particles tend to experience diminished eccentricity variations (see appendix for details), and thus retain low perihelion distances, where they are more likely to be removed by Neptune. Simultaneously, we note that for the exact same reason, such objects are more readily discoverable by astronomical surveys, and are thus bound to be over-represented within the observational census of long-period KBOs. Accordingly, further characterization of P9-sculpted orbital distribution, fully accounting for the overlaying observational biases using high-resolution numerical experiments constitutes an important step towards continued evaluation of the Planet Nine hypothesis within the framework of the emergent dataset.
 
\section{Discussion}\label{sect6}

Within the current observational census of trans-Neptunian objects, the longest-period orbits exhibit unexpected collective structure that is most readily attributed to gravitational perturbations induced by a yet-unseen, massive planet. While numerical simulations aimed at reproducing the Kuiper belt's orbital makeup through gravitational interactions with Planet Nine are now plentiful in the literature \citep{BB16,BroBat,ML2017,Becker2017}, the physics of the dynamical processes responsible for shaping the distant Kuiper belt remains largely unclear \citep{Beust2016}. In this work, we have sought to resolve this problem, and characterize the dynamical evolution induced by Planet Nine upon long-period Kuiper belt objects, from semi-analytic grounds. 

The specific aim of this work has been to qualitatively understand the three primary lines of evidence for the existence of Planet Nine. They are: (i) orbital clustering of long-period Kuiper belt objects, (ii) dynamical detachment of KBO orbits from Neptune, and (iii) generation of highly inclined/retrograde bodies within the solar system. We note that none of these effects are new, and have already been pointed out in the work of \citep{BB16}. Accordingly, the primary purpose of this study has been to create an analytical guide for the interpretation of existing and future numerical results, rather than to generate new ones. In doing so, we have been guided by a series of queries that we outlined in the introduction. Let us now recall, and provide the answers to these questions.

\begin{itemize}
\setlength\itemsep{-0em}
\item What role (if any) do resonant interactions play within the dynamical evolution induced by Planet Nine? If resonances are prevalent, what order/multiplet harmonics dominate the dynamics, and what are their characteristic widths?
\end{itemize}

In the most idealized case of strictly planar physical setup, all high-eccentricity particles that occupy stable orbits within Planet Nine's gravitational domain of influence derive their prolonged dynamical lifetimes from the phase-protection mechanism inherent to mean-motion resonances\footnote{Recall from section \ref{sect2} that stable orbits outside MMRs also exist, and avoid close encounters with P9 via low-amplitude apsidal libration around $\Delta\varpi=0$. Such orbits, however, never reach low values of $q$ and are therefore difficult to detect.}. The critical angles associated with these resonances typically have the form $\phires = \p\,\lambda_9-\q\,\lambda-(\p-\q)\,\varpi_9$, and due to the absence of $\varpi$ (i.e., longitude of perihelion of the KBO) from the expression, these harmonics only modulate the mean anomalies and semi-major axes of the KBOs\footnote{This is evident from simple application of Hamilton's equations to the disturbing potential.}. The most prevalent interior resonances that arise in our calculations correspond to the 1:1, 3:2, 2:1 and 5:2 period ratios, with secondary contributions from the 5:4, 4:3, 5:3, and 3:1 commensurabilities. Provided nominal Planet Nine parameters, the widths of these resonances lie on the order of $\delta a\sim 2-15\,$AU, and scale as $\delta a\propto m_9^{1/2}$ \citep{1983CeMec..30..197H}.

If the strict coplanarity restriction is lifted, and the particles are endowed with a small inclination dispersion relative to the plane of Planet Nine's orbit (as in the real Kuiper belt), a large portion of resonant dynamics become chaotic, and facilitate an essentially diffusive semi-major axis evolution, with particles hopping from one resonance to another. Moreover, for high eccentricity orbits that reach $q\lesssim36\,$AU, stochastic semi-major axis transport is further enhanced by gravitational scattering off of Neptune \citep{2008ssbn.book..259G}. These complications imply that only dynamically detached objects with $q\gtrsim40\,$AU can reasonably be speculated to currently reside in mean-motion resonances with Planet Nine, and any attempt to calculate the present-day semi-major axis of Planet Nine exclusively from resonant relationships with the observed KBOs \citep{Malhotra2016} may be spoiled by the chaotic nature of the underlying dynamics.

\begin{itemize}
\setlength\itemsep{-0em}
\item What role (if any) do secular interactions play within the dynamical evolution induced by Planet Nine? If dominant, how are close encounters avoided on co-planar, anti-aligned orbits? Moreover, if resonant interactions are relevant to the Planet Nine hypothesis, why does the purely secular phase-space portrait provide a good match to the results of numerical simulations?
\end{itemize}

While mean-motion resonances stabilize the orbits against close-encounters, they do little to modulate the eccentricities and longitudes of perihelia of the affected bodies. As a result, the $e-\Delta\varpi$ dynamics induced upon distant KBOs by Planet Nine is largely secular. This explains why the doubly averaged treatment of the dynamics outlined in section \ref{sect2} provides a good approximation to the results of numerical simulations discussed in section \ref{sect3} \citep{Beust2016}. At the same time, it is important to keep in mind that the true resonant-secular evolution facilitated by Planet Nine's mean gravitational potential is subtly different from the purely secular limit, since the orbital averaging process itself is subject to the resonant relationship among the orbital phases of the bodies. The delicate differences between purely secular and resonant-secular dynamics induced by Planet Nine can be noted by comparing Figures (\ref{Fig:secular}) and (\ref{Fig:secres}).

The eccentricity-perihelion projection of the resonant-secular phase-space portraits outlined in Figure (\ref{Fig:secres}) further shows that apsidal clustering of distant KBOs and the detachment of perihelion from Neptune's orbit exemplified by objects such as Sedna and 2012$\,$VP$_{113}$ \citep{2004ApJ...617..645B,TS14} are actually the same physical effect. That is, as the orbits of KBOs evolve along the level curves of the resonant-secular Hamiltonian, they are forced to encircle a stable equilibrium that resides at $\Delta\varpi=180\deg$. Thus, libration of a Kuiper belt object's longitude of perihelion around the apsidally anti-aligned configuration with respect to Planet Nine is accompanied by conjugate oscillations of the eccentricity that periodically detach (and reattach) the perihelion from (to) Neptune. 

While the gravitational influence of P9 alone provides a perfectly adequate mechanism for perihelion detachment of long-period objects, we cannot exclude the possibility that KBOs were additionally affected by other dynamical processes (such as interactions with the birth cluster; \citealt{2004AJ....128.2564M}) during the solar system's infancy. If so, some fraction of the distant clustered population may occupy secular cycles that will never bring their perihelia sufficiently close to the orbit of Neptune for scattering to ensue. In either case, our calculations reveal that the maximal width of the perihelion cluster is limited to $| \Delta \varpi - 180\deg |\lesssim 90\deg$ - a restriction facilitated by both the character of the dynamics itself, as well as the collision locus that limits the domain where close encounters can be avoided on the $e-\Delta\varpi$ plane. Within this framework, Planet Nine's mass merely regulates the size of the chaotic layer, and the timescale on which perihelion cluster get sculpted.

\begin{itemize}
\setlength\itemsep{-0em}
\item What parameters determine the critical semi-major axis corresponding to the transition between randomized and clustered longitudes of perihelion? What physical effect controls this transition?
\end{itemize}

Given that the long-term dynamics induced upon KBOs by Planet Nine are essentially secular in nature, the critical semi-major axis beyond which orbital clustering ensues corresponds to a point where quadrupolar torques induced by Planet Nine begin to dominate over those arising from the canonical giant planets. Computed in this framework, curves corresponding to critical semi-major axes of $a_{\rm{crit}} = 150, 200$ and $250\,$AU are delineated in $e_9-a_9$ space for $m_9=5,10,$ and $20\,m_\oplus$ in Figure (\ref{Fig:acrit}). While these loci provide an approximate measure of $a_{\rm{crit}}$, we note that the transition between apsidally confined and randomized orbits in our $N$-body simulations is somewhat gradual (as shown in Figure \ref{Fig:histogram}). 

With decreasing semi-major axis, the relative number of resonant bodies that are apsidally clustered decreases with respect to those that are not\footnote{In our coplanar calculations, this transition also corresponds to the progressive dominance of the angle $\psires$ over $\phires$ (see equation \ref{phi}) as the resonant guiding center.}. This means that although $a_{\rm{crit}}$ provides a characteristic semi-major axis that corresponds to the onset of orbital clustering, the real Kuiper belt at $a \lesssim a_{\rm{crit}}$ will show orbital clustering that is increasingly contaminated by non-anti-aligned bodies - an effect seen in the real data \citep{TS14,Shankman}. Meanwhile, our calculations suggest that clustering of the orbital planes of distant KBOs and the corresponding nodal confinement is a simple consequence of the tilting of the Laplace plane away from the ecliptic by Planet Nine. 

\begin{itemize}
\setlength\itemsep{-0em}
\item What is the qualitative behavior of inclination dynamics within the framework of P9-driven evolution? What dynamical process allows some of the objects to acquire exceptionally high inclinations in the distant Kuiper belt?
\end{itemize}

Not all objects affected by Planet Nine remain confined to the Laplace plane on multi-Gyr timescales. Instead, a subset of long-period KBOs execute large-scale oscillations in the orbital inclination as well as eccentricity. This mode of P9-induced evolution is distinct from the Kozai-Lidov mechanism, and is driven by a high-order secular resonance that is characterized by simultaneous libration of the critical angle $\theta=\Delta\varpi-2\omega$ as well as $\Delta\varpi$. This doubly-resonant form of secular coupling forces a particular strenuous exchange of angular momentum, and generally leads to acute orbit-flipping behavior of distant KBOs \citep{2014ApJ...785..116L}.

The onset of these large-amplitude orbital excursion is fundamentally chaotic, and is facilitated by variations in the the angular momentum deficit that is driven by resonant-secular $e-\Delta\varpi$ oscillations. That is, oscillations in the eccentricity modulate the system's proximity to the $\theta-$resonance, periodically allowing low-inclination orbits to enter the highly inclined dynamical regime. As the subsequent orbital evolution unfolds, the eccentricity reaches a peak of its cycle when inclination is approximately $i\approx 90 \deg$, meaning that bodies belonging to the highly inclined population are most readily observable in a state roughly perpendicular to the plane of the solar system. This qualitatively explains the observed dynamical state of large semi-major axis centaurs, which constitutes the third major line of evidence for the existence of Planet Nine. \\

There exist two other, secondary lines of evidence for the existence of Planet Nine, which we did not discuss in this paper. The first is obliquity of the Sun \citep{2016AJ....152..126B,2016AJ....152..215L,2017AJ....153...27G}. One reason we chose to not discuss this dynamical effect is because it is intrinsically trivial. It is well-known that secular coupling between two mutually inclined orbits results in a regression of the node of both orbits, meaning that the twisting of the giant planets' orbital plane out of alignment with the solar spin-axis is an inescapable consequence of Planet Nine's existence. As a result, the genuinely remarkable aspect of this calculation is not that a spin-orbit misalignment can be excited, but the fact that a Planet Nine configuration close to the one deduced from distant Kuiper belt constraints can adequately reproduce the solar obliquity, when its gravitational influence is exerted over the entire lifetime of the solar system. 

At the same time, we note that strictly speaking, Planet Nine is not \textit{required} to explain the obliquity of the sun, and other theoretical models exist. For example, \citet{2012Natur.491..418B} argued that a primordial binary companion to the solar system could have excited the observed spin-orbit misalignment through the same exact mechanism, while \citet{2011MNRAS.412.2790L} have proposed that magnetic interactions between the sun and the inner regions of the protosolar nebula could have accomplished the same task. We note however, that while a multitude of processes could have contributed to the observed obliquity of the sun, the gravitational influence of Planet Nine provides the only dynamical mechanism that is directly testable.

Another population of objects that Planet Nine naturally generates is the highly inclined component of the proximate ($a<100\,$AU) Kuiper belt. This subset of bodies includes all objects with $i\gtrsim35\deg$ that do not naturally emerge from simulations of the solar system's primordial evolution \citep{2008Icar..196..258L,2011ApJ...738...13B,2015AJ....150...73N}, and include the retrograde objects ``Drac" \citep{2009ApJ...697L..91G} and ``Niku" \citep{2016ApJ...827L..24C}. While these objects appear observationally distinct from the nearly-perpendicular high semi-major axis centaurs discussed above, full-fledged $N$-body simulations reported in \citep{BB16b} show that the highly inclined $a<100\,$AU objects are simply the large-$a$ objects that have been scattered inwards by Neptune. As a consequence, they do not require a separate dynamical explanation from the objects undergoing large-scale orbital excursions discussed in section \ref{sect52} of this paper. 

Cumulatively, the aforementioned lines of evidence constitute a compelling case for the existence of Planet Nine, as none of the objects within the current observational dataset exert any significant tension upon the model. In light of this broadranging agreement, (short of direct detection of Planet Nine) the most direct avenue towards further reinforcement or falsification of our theory is the continued detection of $a\gtrsim250\,$AU Kuiper belt objects with the aim to better establish the statistical significance of the clustering of longitudes of perihelion and ascending node \citep{2017AJ....154...65B,Shankman}. To this end, we reiterate that even though contamination of the clustered orbital pattern by unconfined particles is an expected result of P9's gravitational influence, a notable grouping of long-period trajectories in physical space remains a key feature of the dynamical model.

While the evidence for P9 remains strong, as already discussed in the introduction, a simple proposition of an extant planet beyond Neptune does not amount to a meaningful theoretical prediction. Instead, the Planet Nine hypothesis is uniquely defined by the combination of observational signatures it explains and the specific dynamical mechanisms through which these astronomical patterns arise. Thus, the final evaluation of the Planet Nine hypothesis will not simply correspond to a detection of a planet beyond Neptune, but the confrontation of the outlined theory with the dynamical evolution induced by this planet. Thankfully, the observational prospects for the direct detection of Planet Nine either through ongoing or future surveys are quite promising \citep{BroBat,2016AJ....152...94H,2016AJ....152...80H,2016ApJ...824L..25F,ML2017}, and it is likely that the concluding assessment of the theoretical model outlined in this paper will occur on a timescale considerably shorter than a decade.

\acknowledgments
\textbf{Acknowledgments}. We are thankful to Mike Brown, Greg Laughlin, Chris Spalding, Matt Holman, Gongjie Li, Tali Khain and Elizabeth Bailey for illuminating discussions. Additionally, we are grateful to Sarah Millholland for providing a thorough and insightful referee report, as well as to Charles Fairchild, whose generous support facilitated this collaboration. 

\begin{appendix}

\section{Computational Details}

Throughout the paper, we have relied on various closed-form computations of the averaged Hamiltonian (specifically, expressions \ref{Hpuresec}, \ref{Hres} and \ref{Hsecres}), without explicitly stating how the calculations were carried out. Let us now comment on the practical details inherent to these evaluations. In the plane, the cartesian coordinates of the particle's position vector, $\textbf{r}$, are given by the well-known relations \citep{Morby2002}:
\begin{align}
x=a\big(\cos E-e \big)\cos \varpi-a\sqrt{1-e^2}\sin E\sin \varpi \nonumber \\
y=a\big(\cos E-e \big)\sin \varpi-a\sqrt{1-e^2}\sin E\cos \varpi,
\label{cartvar}
\end{align}
where $E$ is the eccentric anomaly. Identical expressions (with subscript 9) apply to the position vector of Planet Nine.

While the aforementioned cartesian coordinates are most naturally expressed in terms of the eccentric anomaly, canonical averaging of the Hamiltonian is carried out with respect to the mean longitude, $\lambda$. The two quantities are related via Kepler's equation. Therefore, rather than solving Kepler's equation at every computational step to express $x$ and $y$ in terms of $\lambda$, it is more convenient to integrate directly with respect to $E$, by introducing the Jacobian \citep{Gabrielle}
\begin{align}
d\lambda=(1-e\,\cos E)\,dE.
\label{jacobian}
\end{align}

For the case of the doubly averaged secular Hamiltonian (\ref{Hpuresec}), it is appropriate to carry out the integral assuming that $\lambda$ and $\lambda_9$ are not correlated. This is however not the case for singly averaged resonant and resonant-secular Hamiltonians (\ref{Hres}) and (\ref{Hsecres}). Particularly, in these instances the mean longitudes of the particle and Planet Nine are linked to one-another through the resonant relationship:
\begin{align}
\lambda=\frac{\phires-(\p-\q)\,\varpi_9+\p\,\lambda_9}{\q}.
\label{resrel}
\end{align}
Accordingly, in this case Kepler's equation must be solved at every iteration, to obtain the eccentric anomaly of the particle as a function of $\lambda_9$.

In order to construct resonant-secular phase-space portraits in the $\mathcal{J}\rightarrow0$ limit, it is necessary to first map out the equilibrium value of the resonant angle $\phires$ on the $(e,\Delta\varpi)$ domain. For the specific problem at hand, this can be done by sampling the resonant Hamiltonian (\ref{Hres}) as a function of $\phires$ along the $a=(\ell/k)^{2/3}\,a_9$ line, and numerically finding the the relevant local maximum. Incidentally, the equilibrium value of $\phires$ typically lies half-way in between the collision curves, which are easily obtained by solving the simultaneous equations $x=x_9,y=y_9$, and substituting the resulting values of $E$ and $E_9$ into Kepler's equation to yield the values of $\phires$ that correspond to collisional trajectories. 

\section{Apsidally Circulating Resonant Orbits}

In section \ref{sect4} of the main text, we constructed a semi-analytic model for apsidally confined resonant orbits residing in interior 5:2, 2:1, 3:2 and 1:1 mean motion resonances with P9. As demonstrated in Figure (\ref{Fig:histogram}), however, particles residing at somewhat lower values of semi-major axes (particularly those entrained in the 3:1 and 4:1 commensurabilities) predominantly exhibit apsidal circulation. Let us briefly consider the dynamical behavior of such trajectories in greater detail.

\begin{figure*}[t]
\centering
\includegraphics[width=1\textwidth]{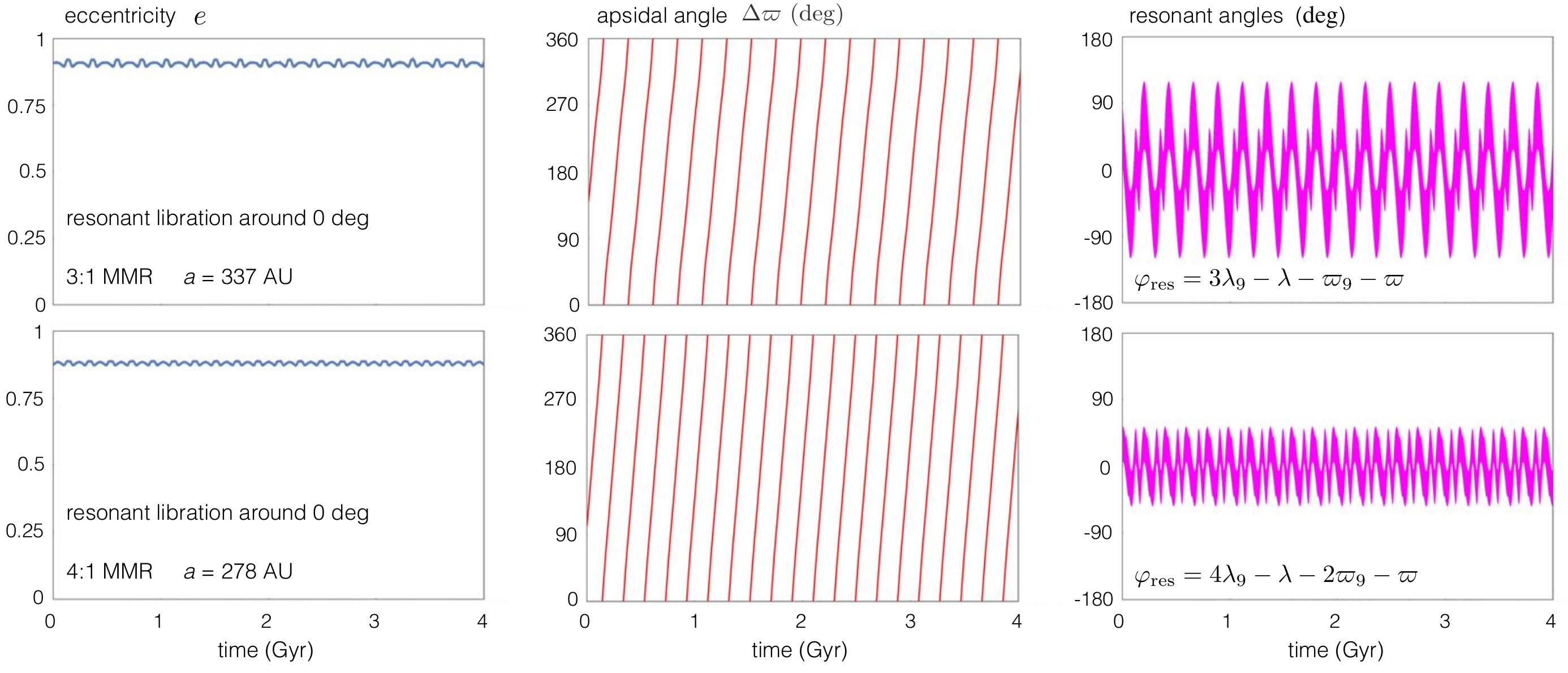}
\caption{Orbital time series of two resonant objects exhibiting apsidal circulation. The trajectories are drawn from the same simulation as those depicted in Figure (\ref{Fig:timeseries}), and correspond to the 3:1 (top) and 4:1 (bottom) resonances respectively. Unlike the apsidally confined orbits depicted in Figure (\ref{Fig:timeseries}), these objects derive their resonant phase-protection mechanism from libration of resonant angles that contain the particle's longitude of perihelion, $\varpi$ (equation \ref{varphi}).}
\label{Fig:circulators}
\end{figure*}

Representative trajectories of particles locked in 3:1 and 4:1 resonances, drawn from the $a=700\,$AU simulation described in section (\ref{sect3}), are shown in Figure (\ref{Fig:circulators}). The panels in this Figure are analogous to those depicted in Figure (\ref{Fig:timeseries}) of the main text, with the exception that the librating resonant angle which drives the dynamics has the form
\begin{align}
\varres &= \p\,\lambda_9-\q\,\lambda-\varpi-(\p-\q-1)\,\varpi_9=\phires-\Delta\varpi.
\label{varphi}
\end{align}
Because these orbits are characterized by circulation of $\Delta\varpi$, no resonant multiplets other than $\varres$ are in libration. Note further that eccentricity modulation associated with the evolution of the apsidal angle $\Delta\varpi$ is in this case very mild, especially compared to that shown in Figure (\ref{Fig:timeseries}).

In order to more closely examine the phase-space evolution of the apsidally circulating orbits, we follow the same semi-analytic procedure as that outlined in section \ref{sect4}. To avoid redundancy, we focus exclusively on the 3:1 MMR, since the 4:1 MMR exhibits very similar behavior. In similitude with equation (\ref{transvar}), we define the canonical coordinates that identify $\varres$ as a reference angle
\begin{align}
&\Lambda''= \Lambda/\q & \varres''=-\varres \nonumber \\
&\Lambda''_9= \Lambda_9+(\p/\q)\, \Lambda  &\lambda''_9=\lambda_9  \nonumber \\
&\Gamma''=\Gamma+\Lambda/\q &\Delta \gamma=\gamma-\gamma_9 = -\Delta\varpi \nonumber \\
&\Gamma''_9=\Gamma_9+\Gamma+((\p-\q)/\q)\,\Lambda  &\gamma''_9=\gamma_9.
\label{transvar2}
\end{align}
As before, upon averaging the Hamiltonian with respect to the fast angle $\lambda''_9$, we are left behind with an adiabatic system that is characterized by a resonant degree of freedom in ($\Lambda''-\varres''$) as well as a secular degree of freedom in ($\Gamma''-\Delta \gamma$).

\begin{figure*}[t]
\centering
\includegraphics[width=0.9\textwidth]{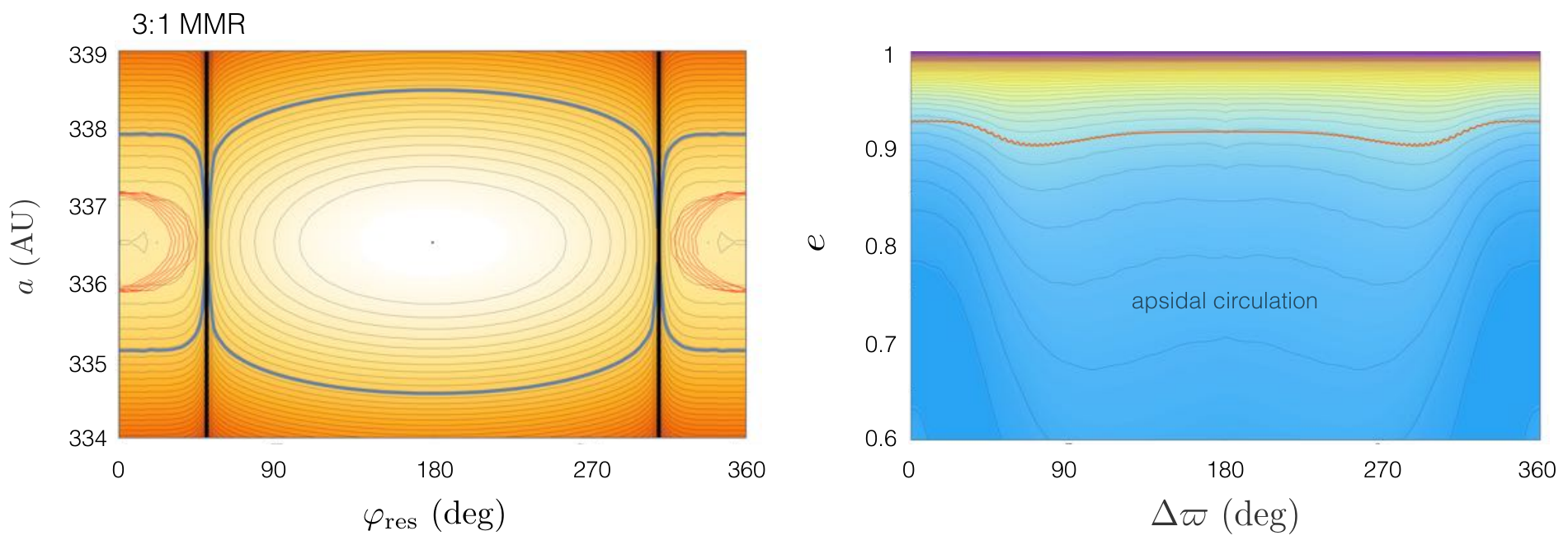}
\caption{Semi-analytically computed resonant $a-\varphi$ (left) and resonant-secular $e-\Delta\varpi$ (right) phase-space portraits for the 3:1 MMR, assuming equilibrium libration of the critical angle given by expression (\ref{varphi}). Numerically computed trajectories shown in Figure (\ref{Fig:circulators}) are depicted with red and orange lines in the left and right panel respectively, signaling satisfactory agreement between semi-analytic theory and $N$-body simulations. See captions of Figures (\ref{Fig:resonant}) and (\ref{Fig:secres}) for further details.}
\label{Fig:circan}
\end{figure*}

Freezing $\Gamma''$ at a value that corresponds to $q=35\,$AU at $a=(\ell/k)^{2/3}\,a_9$ and setting $\Delta \varpi=-\Delta \gamma=\pi$, we have computed the resonant phase-space diagram akin to those presented in Figure (\ref{Fig:resonant}) for the interior 3:1 resonance. However, in this case, the Hamiltonian (\ref{Hres}) was computed under the constraint of the resonant relationship (\ref{varphi}) instead of equation (\ref{phi}). The resulting ($a-\varres$) diagram is shown in the left panel of Figure (\ref{Fig:circan}) with the result of our $N$-body simulation over-plotted in red.

Unlike the long-term stable orbits that exhibit steady perihelion clustering, apsidally circulating trajectories shown in Figure (\ref{Fig:circulators}) reside in the sub-domain of the ($a-\varres$) diagram that is occupied by the $\infty$-shaped separatrix, and encircle this homoclinic curve from the outside. In contrast with the neighborhood of the elliptic equilibrium point discussed in section \ref{sect41}, this phase-space domain is never fully swept by the collision curves as $\Delta\varpi$ swings from $0$ to $2\pi$. As a result, trajectories that surround this resonance can safely rotate through all possible values of $\Delta\varpi$ without compromising the resonant phase-protection mechanism.

This means that in the case of resonances characterized by libration of $\varres$, there is no equivalent of the $(e-\Delta\varpi)$ collision locus that arises within the framework of MMRs characterized by libration of $\phires$ (shown as the bounding curves in Figure \ref{Fig:secres}). Nevertheless, the resonance that resides at the core of the apsidally circulating trajectory (as shown in Figure \ref{Fig:circan}) undergoes topological changes driven by evolution of $\Delta\varpi$. In particular, as $\Delta\varpi$ shifts away from $\pi$ towards $0$, the $\infty$-shaped separatrix deforms asymmetrically and eventually vanishes, as two of the three associated fixed points disappear. Finally, as $\Delta\varpi$ tends closer to $0$, the remaining (elliptic) fixed point returns to the origin, such that the trajectory once again encircles $\varres=0$ at $\Delta\varpi=0$\footnote{The same picture applies to $\Delta\varpi$ shifting from $\pi$ to $2\pi$, but the opposite resonant equilibrium point survives in this case.}. In other words, as $\Delta\varpi$ circulates from $0$ to $2\pi$, the resonant center of $\varres$ experiences a concurrent oscillation about $\varres=0$ with an amplitude of order $\sim\pi/2$ (as shown in the right panel of Figure \ref{Fig:circulators}).


In order to approximately elucidate the resonant-secular dynamics that unfolds in the $\Delta\varpi$-circulating regime, we have computed the averaged Hamiltonian (\ref{Hsecres}) under the constraint of the resonant relationship (\ref{varphi}). Following the discussion outlined in section (\ref{sect42}), we assume that the adiabatic invariant $\mathcal{J}=0$, and thereby confine ourselves to the $a-\varres$ equilibrium point that remains extant (and stable) in the $\Delta\varpi\in(0,\pi)$ range, and simply reflect the computed portrait onto the $\Delta\varpi\in(\pi,2\pi)$ domain. The corresponding $(e-\Delta\varpi)$ diagram is shown in the right panel of Figure (\ref{Fig:circan}), with the $N$-body trajectory over-plotted in orange. 

Although the semi-analytical secular portrait matches the numerically computed evolution well, we note that for these apsidally circulating trajectories, the $\mathcal{J}=0$ assumption is a relatively crude one, since the orbit itself resides outside the separatrix and does not directly encircle the equilibrium point. To this end, the kinks in the semi-analytical curves shown in Figure (\ref{Fig:circan}) at $\Delta\varpi=\pi$ are an unphysical consequence of this assumption, and would disappear if the portrait was more carefully computed by carrying out the averaging process along the contour of a ($a-\varres$) trajectory that encircles a phase-space area greater than that occupied by the separatrix. In fact, the reason for which the $\mathcal{J}=0$ approximation works relatively well in our calculation is that the area engulfed by the $\infty$-shaped curve is never big. Therefore, a more rigorous treatment of adiabatic theory can be carried out by assuming a non-zero but nevertheless small value of $\mathcal{J}$. 

Irrespective of the details associated with computation of the Hamiltonian (\ref{Hsecres}) for $\varres$-type commensurabilities, the results of the $N$-body simulation shown in Figure (\ref{Fig:circulators}) as well as the semi-analytic calculations presented in Figure (\ref{Fig:circan}), point to the fact that resonantly protected trajectories that circulate in longitude of perihelion naturally arise within the framework of the Planet Nine hypothesis. Correspondingly, given sufficient data, characterization of the specific semi-major axes that such trajectories occupy may provide key constraints on the present-day orbital state of Planet Nine.

\end{appendix}


\begin{thebibliography} 


\bibitem[Adams(1846)]{Adams1846} Adams, J.~C.\ 1846, \mnras, 7, 149 

\bibitem[Adams(2010)]{2010ARA&A..48...47A} Adams, F.~C.\ 2010, \araa, 48, 47 



\bibitem[Bailey et al.(2016)]{2016AJ....152..126B} Bailey, E., Batygin, K., \& Brown, M.~E.\ 2016, \aj, 152, 126 

\bibitem[Bailey et al.(2017)]{Baileyinprep} Bailey, E., Batygin, K., \& Brown, M.~E.\ 2017, in prep.

\bibitem[Batygin et al.(2011)]{2011ApJ...738...13B} Batygin, K., Brown, M.~E., \& Fraser, W.~C.\ 2011, \apj, 738, 13 


\bibitem[Batygin(2012)]{2012Natur.491..418B} Batygin, K.\ 2012, \nat, 491, 418 

\bibitem[Batygin \& Brown(2016a)]{BB16} Batygin, K., \& Brown, M.~E.\ 2016a, \aj, 151, 22

\bibitem[Batygin \& Brown(2016b)]{BB16b} Batygin, K., \& Brown, M.~E.\ 2016b, \apjl, 833, L3 

\bibitem[Beaug{\'e} et al.(2006)]{Beauge2006} Beaug{\'e}, C., Michtchenko, T.~A., \& Ferraz-Mello, S.\ 2006, \mnras, 365, 1160 

\bibitem[Becker et al.(2017)]{Becker2017} Becker, J.~C., Adams, F.~C., Khain, T., Hamilton, S.~J., \& Gerdes, D.\ 2017, \aj, 154, 61 

\bibitem[Beust(2016)]{Beust2016} Beust, H.\ 2016, \aap, 590, L2 

\bibitem[Brown et al.(2004)]{2004ApJ...617..645B} Brown, M.~E., Trujillo, C., \& Rabinowitz, D.\ 2004, \apj, 617, 645 

\bibitem[Brown \& Batygin(2016)]{BroBat} Brown, M.~E., \& Batygin, K.\ 2016, \apjl, 824, L23 

\bibitem[Brown(2017)]{2017AJ....154...65B} Brown, M.~E.\ 2017, \aj, 154, 65 


\bibitem[Brunini \& Melita(2002)]{2002Icar..160...32B} Brunini, A., \& Melita, M.~D.\ 2002, \icarus, 160, 32 


\bibitem[Chambers(1999)]{1999MNRAS.304..793C} Chambers, J.~E.\ 1999, \mnras, 304, 793 

\bibitem[Chen et al.(2016)]{2016ApJ...827L..24C} Chen, Y.-T., Lin, H.~W., Holman, M.~J., et al.\ 2016, \apjl, 827, L24 



\bibitem[Delisle et al.(2012)]{2012A&A...546A..71D} Delisle, J.-B., Laskar, J., Correia, A.~C.~M., \& Bou{\'e}, G.\ 2012, \aap, 546, A71 




\bibitem[Forbes(1880)]{Forbes1880} Forbes, G.\ 1880, \nat, 21, 562 

\bibitem[Fortney et al.(2016)]{2016ApJ...824L..25F} Fortney, J.~J., Marley, M.~S., Laughlin, G., et al.\ 2016, \apjl, 824, L25 



\bibitem[Gladman et al.(2002)]{2002Icar..157..269G} Gladman, B., Holman, M., Grav, T., et al.\ 2002, \icarus, 157, 269 

\bibitem[Gladman \& Chan(2006)]{2006ApJ...643L.135G} Gladman, B., \& Chan, C.\ 2006, \apjl, 643, L135 

\bibitem[Gladman et al.(2009)]{2009ApJ...697L..91G} Gladman, B., Kavelaars, J., Petit, J.-M., et al.\ 2009, \apjl, 697, L91 


\bibitem[Gomes et al.(2006)]{2006Icar..184..589G} Gomes, R.~S., Matese, J.~J., \& Lissauer, J.~J.\ 2006, \icarus, 184, 589 

\bibitem[Gomes et al.(2008)]{2008ssbn.book..259G} Gomes, R.~S., Fern Ndez, J.~A., Gallardo, T., \& Brunini, A.\ 2008, The Solar System Beyond Neptune, 259 

\bibitem[Gomes et al.(2015)]{Gomes2015} Gomes, R.~S., Soares, J.~S., \& Brasser, R.\ 2015, \icarus, 258, 37 

\bibitem[Gomes et al.(2017)]{2017AJ....153...27G} Gomes, R., Deienno, R., \& Morbidelli, A.\ 2017, \aj, 153, 27 

\bibitem[Gronchi \& Milani(1998)]{Gronchi1998} Gronchi, G.~F., \& Milani, A.\ 1998, Celestial Mechanics and Dynamical Astronomy, 71, 109 

\bibitem[Gronchi(2002)]{2002CeMDA..83...97G} Gronchi, G.~F.\ 2002, Celestial Mechanics and Dynamical Astronomy, 83, 97 



\bibitem[Henrard \& Lamaitre(1983)]{1983CeMec..30..197H} Henrard, J., \& Lamaitre, A.\ 1983, Celestial Mechanics, 30, 197 


\bibitem[Henrard \& Caranicolas(1990)]{1990CeMDA..47...99H} Henrard, J., \& Caranicolas, N.~D.\ 1990, Celestial Mechanics and Dynamical Astronomy, 47, 99 

\bibitem[Henrard(1993)]{Henrard1993} Henrard J., 1993, The Adiabatic Invariant in Classical Mechanics, Dynamics Reported: New Series, Vol. 2, New York: Springer

\bibitem[Holman \& Payne(2016a)]{2016AJ....152...94H} Holman, M.~J., \& Payne, M.~J.\ 2016a, \aj, 152, 94 

\bibitem[Holman \& Payne(2016b)]{2016AJ....152...80H} Holman, M.~J., \& Payne, M.~J.\ 2016b, \aj, 152, 80 


\bibitem[Hoyt(1980)]{Hoyt1980} Hoyt, W.~G.\ 1980, Planets X and Pluto, Tucson, AZ: University of Arizona Press 




\bibitem[Kaula(1962)]{1962AJ.....67..300K} Kaula, W.~M.\ 1962, \aj, 67, 300 

\bibitem[Kinoshita \& Nakai(1999)]{1999CeMDA..75..125K} Kinoshita, H., \& Nakai, H.\ 1999, Celestial Mechanics and Dynamical Astronomy, 75, 125 

\bibitem[Kozai(1962)]{1962AJ.....67..591K} Kozai, Y.\ 1962, \aj, 67, 591 



\bibitem[Lai et al.(2011)]{2011MNRAS.412.2790L} Lai, D., Foucart, F., \& Lin, D.~N.~C.\ 2011, \mnras, 412, 2790 

\bibitem[Lai(2016)]{2016AJ....152..215L} Lai, D.\ 2016, \aj, 152, 215 

\bibitem[Laplace(1799)]{Laplace} Marquis de Laplace, P.-S.\ 1799, Trait\'e de m\'ecanique c\'eleste, Paris: Duprat

\bibitem[Lee(2004)]{2004ApJ...611..517L} Lee, M.~H.\ 2004, \apj, 611, 517 


\bibitem[LeVerrier(1846)]{LeVerrier1846} Le Verrier, U.~J. J.\ 1846, Comptes Rendus, 23, 428 

\bibitem[Levison et al.(2008)]{2008Icar..196..258L} Levison, H.~F., Morbidelli, A., Van Laerhoven, C., Gomes, R., \& Tsiganis, K.\ 2008, \icarus, 196, 258 

\bibitem[Li et al.(2014)]{2014ApJ...785..116L} Li, G., Naoz, S., Kocsis, B., \& Loeb, A.\ 2014, \apj, 785, 116 


\bibitem[Lidov(1962)]{1962P&SS....9..719L} Lidov, M.~L.\ 1962, \planss, 9, 719 


\bibitem[Lowell(1915)]{1915MmLow...1....1L} Lowell, P.\ 1915, Memoir on a Trans-Neptunian Planet,  Memoirs of the Lowell Observatory.~John Wiley \& Sons

\bibitem[Lykawka \& Mukai(2008)]{2008AJ....135.1161L} Lykawka, P.~S., \& Mukai, T.\ 2008, \aj, 135, 1161 


\bibitem[Malhotra et al.(2016)]{Malhotra2016} Malhotra, R., Volk, K., \& Wang, X.\ 2016, \apjl, 824, L22 

\bibitem[Mardling(2010)]{2010MNRAS.407.1048M} Mardling, R.~A.\ 2010, \mnras, 407, 1048 


\bibitem[Mardling(2013)]{2013MNRAS.435.2187M} Mardling, R.~A.\ 2013, \mnras, 435, 2187 


\bibitem[Matese \& Whitmire(1986)]{Whitmire1984} Matese, J.~J., \& Whitmire, D.~P.\ 1986, \icarus, 65, 37 

\bibitem[Millholland \& Laughlin(2017)]{ML2017} Millholland, S., \& Laughlin, G.\ 2017, \aj, 153, 91 

\bibitem[Michtchenko et al.(2006)]{Michtchenko2006} Michtchenko, T.~A., Ferraz-Mello, S., \& Beaug{\'e}, C.\ 2006, \icarus, 181, 555 

\bibitem[Morbidelli \& Moons(1993)]{1993Icar..102..316M} Morbidelli, A., \& Moons, M.\ 1993, \icarus, 102, 316 

\bibitem[Morbidelli(2002)]{Morby2002} Morbidelli, A.\ 2002, Modern Celestial Mechanics: Aspects of Solar System Dynamics, London: Taylor \& Francis

\bibitem[Morbidelli \& Levison(2004)]{2004AJ....128.2564M} Morbidelli, A., \& Levison, H.~F.\ 2004, \aj, 128, 2564 

\bibitem[Murray \& Dermott(1999)]{MD99} Murray, C.~D., \& Dermott, S.~F.\ 1999, Solar System Dynamics, Cambridge, UK: Cambridge University Press


\bibitem[Neishtadt(1984)]{1984PriMM..48..197N} Neishtadt, A.~I.\ 1984, Prikladnaia Matematika i Mekhanika, 48, 197 

\bibitem[Nesvorn{\'y}(2015)]{2015AJ....150...73N} Nesvorn{\'y}, D.\ 2015, \aj, 150, 73 

\bibitem[Nesvorny et al.(2017)]{2017arXiv170607447N} Nesvorny, D., Vokrouhlicky, D., Dones, L., et al.\ 2017, arXiv:1706.07447 



\bibitem[Pichierri et al.(2017)]{Gabrielle} Pichierri, G., Morbidelli, A., \& Lai, D.\ 2017, arXiv:1705.01841 


\bibitem[Pickering \& Pickering(1909)]{1909AnHar..61..109P} Pickering, W.~H., \& Pickering, E.~C.\ 1909, Annals of Harvard College Observatory, 61, 109 

\bibitem[Poincar{\'e}(1902)]{Poincare1902} Poincar{\'e}, H.\ 1902, Bulletin Astronomique, Serie I, 19, 289 

\bibitem[Press et al.(1992)]{1992nrca.book.....P} Press, W.~H., Teukolsky, S.~A., Vetterling, W.~T., \& Flannery, B.~P.\ 1992, Numerical recipes in FORTRAN. The art of scientific computing, Cambridge: University Press





\bibitem[Saillenfest et al.(2017)]{2017arXiv170701379S} Saillenfest, M., Fouchard, M., Tommei, G., \& Valsecchi, G.~B.\ 2017, arXiv:1707.01379 

\bibitem[Shankman et al.(2017)]{Shankman} Shankman, C., Kavelaars, J.~J., Bannister, M.~T., et al.\ 2017, \aj, 154, 50 


\bibitem[Tan et al.(2013)]{2013ApJ...777..101T} Tan, X., Payne, M.~J., Lee, M.~H., et al.\ 2013, \apj, 777, 101 

\bibitem[Thomas \& Morbidelli(1996)]{1996CeMDA..64..209T} Thomas, F., \& Morbidelli, A.\ 1996, Celestial Mechanics and Dynamical Astronomy, 64, 209 

\bibitem[Trujillo \& Sheppard(2014)]{TS14} Trujillo, C.~A., \& Sheppard, S.~S.\ 2014, \nat, 507, 471 




\bibitem[Volk \& Malhotra(2017)]{2017AJ....154...62V} Volk, K., \& Malhotra, R.\ 2017, \aj, 154, 62 



\bibitem[Wisdom(1985)]{1985Icar...63..272W} Wisdom, J.\ 1985, \icarus, 63, 272 

\bibitem[Wisdom \& Holman(1992)]{1992AJ....104.2022W} Wisdom, J., \& Holman, M.\ 1992, \aj, 104, 2022 







\end{thebibliography}
\end{document}